\documentclass[fleqn,usenatbib]{mnras}

\usepackage{newtxtext,newtxmath}
\usepackage{amsmath}
\usepackage[T1]{fontenc}
\usepackage{ae,aecompl}

\usepackage{appendix}
\usepackage{graphicx}% Include figure files
\usepackage{xcolor}
\usepackage{dcolumn}% Align table columns on decimal point
\usepackage{bm}% bold math
\usepackage{aas_macros}
\usepackage{cleveref}	% adds cref command
\usepackage{enumitem}
\usepackage{csquotes}
\usepackage[all]{nowidow}		% tries to avoid widows
\usepackage{placeins}

\graphicspath{{figs/}}
\title[Growth of supermassive black holes from the collapse of dark matter cores]{On the growth of supermassive black holes formed from the gravitational collapse of fermionic dark matter cores}

\author[C. R. Argüelles et. al.]{C. R. Arg\"uelles,$^{1,2,3}$\thanks{Email: carguelles@fcaglp.unlp.edu.ar} 
K. Boshkayev,$^{4,5}$\thanks{Email: kuantay@mail.ru} 
A. Krut,$^{2}$ 
G. Nurbakhyt,$^{4}$ 
J. A. Rueda,$^{2,3,6,7,8}$\thanks{Email: jorge.rueda@icra.it}
\newauthor
R. Ruffini,$^{2,3,9}$\thanks{Email: ruffini@icra.it} 
J. D. Uribe-Suárez,$^{2,10}$\thanks{Email: juan.uribe@icranet.org} 
and R. Yunis$^{2}$
\\
$^{1}$Instituto de Astrof\'isica de La Plata, UNLP-CONICET, Paseo del Bosque s/n B1900FWA La Plata, Argentina\\
$^{2}$ICRANet, Piazza della Repubblica 10, I-65122 Pescara, Italy\\
$^{3}$ICRA, Dipartimento di Fisica, Sapienza Universit\`a  di Roma, Piazzale Aldo Moro 5, I-00185 Roma, Italy\\
$^{4}$NNLOT, Department of Theoretical and Nuclear Physics, Al-Farabi Kazakh National University, Almaty 050040, Kazakhstan\\
$^{5}$International University of Information Technology, Manas St. 34/1, 050040 Almaty, Kazakhstan.\\
$^{6}$ICRANet-Ferrara, Dipartimento di Fisica e Scienze della Terra, Universit\`a degli Studi di Ferrara, Via Saragat 1, I-44122 Ferrara, Italy\\
$^{7}$Dipartimento di Fisica e Scienze della Terra, Universit\`a degli Studi di Ferrara, Via Saragat 1, I-44122 Ferrara, Italy\\
$^{8}$INAF, Istituto di Astrofisica e Planetologia Spaziali, Via Fosso del Cavaliere 100, I-00133 Rome, Italy\\
$^{9}$INAF, Viale del Parco Mellini 84, I-00136 Rome, Italy\\
$^{10}$Facultad de Ciencias Básicas, Universidad Santiago de Cali, Campus Pampalinda, Calle 5 No. 6200, 760035 Santiago de Cali, Colombia\\
}

\date{Accepted XXX. Received YYY; in original form ZZZ}

\begin{document}
\label{firstpage}
\pagerange{\pageref{firstpage}--\pageref{lastpage}}
\maketitle

\pubyear{2023}

\begin{abstract}
Observations support the idea that supermassive black holes (SMBHs) power the emission at the center of active galaxies. However, contrary to stellar-mass BHs, there is a poor understanding of their origin and physical formation channel. In this article, we propose a new process of SMBH formation in the early Universe that is not associated with baryonic matter (massive stars) or primordial cosmology. In this novel approach, SMBH seeds originate from the gravitational collapse of fermionic dense dark matter (DM) cores that arise at the center of DM halos as they form. We show that such a DM formation channel can occur before star formation, leading to heavier BH seeds than standard baryonic channels. The SMBH seeds subsequently grow by accretion. We compute the evolution of the mass and angular momentum of the BH using a geodesic general relativistic disk accretion model. We show that these SMBH seeds grow to $\sim 10^9$--$10^{10} M_\odot$ in the first Gyr of the lifetime of the Universe without invoking unrealistic (or fine-tuned) accretion rates.
\end{abstract}


% \maketitle

%%%%%%%%%%%%%%%%%%%%%%%%%%%%%%%%%%%%%%%%%%%%%%%%%%
%%%%%%%%%%%%%%%%% BODY OF PAPER %%%%%%%%%%%%%%%%%%

%%%%%%%%%%%%%%%%%%%%%%%%%%%%%%%%%%%%%%%%%%%%%%%%%%%%
%%%%%%%%%%%%%%%%%%%%%%%%%%%%%%%%%%%%%%%%%%%%%%%%%%%%
\section{Introduction}\label{sec:I}
%%%%%%%%%%%%%%%%%%%%%%%%%%%%%%%%%%%%%%%%%%%%%%%%%%%%
%%%%%%%%%%%%%%%%%%%%%%%%%%%%%%%%%%%%%%%%%%%%%%%%%%%%

The formation, growth, and nature of the supermassive BHs (SMBHs) residing at the galaxy centers are outstanding problems in astrophysics and cosmology. Important unresolved questions include: how can they grow so large and so fast to be present in the farthest distant quasars \cite{2012Sci...337..544V,2019PASA...36...27W}; what is the nature and mass of BH seeds that grow to form the SMBHs of $\sim 10^8$--$10^9 M_\odot$ in the high-$z$ Universe \cite{2022MNRAS.tmp.1507Z}; and what is the nature of the connection between the total mass of a host galaxy and the mass of its central SMBH \cite{2021NatRP...3..732V}.

Here we propose a new paradigm for the nature and formation of SMBH seeds, which arise from the gravitational collapse of high-density regions of dark matter (DM) at the galaxy centers \cite{2021MNRAS.502.4227A}. We present calculations on the subsequent growth of such BH seeds from an accretion disk around a Kerr BH in a fully general relativistic framework. In this way, we aim here to provide answers to the above three main questions. 

Among the various scenarios analyzed in the literature to explain the origin of SMBHs (see \cite{2020ARA&A..58...27I, 2021NatRP...3..732V,2022NewAR..9401642M} for recent reviews), we can divide them into two main categories according to their formation channel: (I) channels associated with baryonic matter, i.e., gas and stars, and (II) channels associated with early universe cosmology. In this work, we propose a novel, third possible scenario: (III) channels associated with DM. Before motivating this new framework, we highlight the pros and cons of the most studied formation channels. In the case of the baryonic channels (I), we can distinguish among (a) \textit{Population III stars}, and (b) \textit{direct collapse to a BH (DCBH)}. Pop. III stars are physically motivated (though yet hypothetical) stars with an average mass of $\sim 10^2 M_\odot$ 
\cite{2001ApJ...551L..27M, 2016ApJ...824..119H} originated in metal-free clouds hosted in small halos of $\sim 10^6 M_\odot$ at high $z$. Thus, the stellar collapse of such massive stars would lead to a BH seed of $\sim 10^2 M_\odot$ which, under idealistic accretion conditions, would reach a $10^9 M_\odot$ SMBH in the first billion years. However, recent state-of-the-art simulations show that BH seeds of $<10^3 M_\odot$ fail to grow until $10^8 M_\odot$ at $z \sim 6$ because of strong radiative feedback \cite{2022MNRAS.tmp.1507Z}. In the DCBH scenario (b), dense gas clumps at the center of massive halos of $\sim 10^8 M_\odot$ become globally unstable and collapse first to a supermassive star of $\sim 10^4$--$10^5 M_\odot$, which then undergoes core-collapse to a central BH. The newborn BH then grows fast by accreting surrounding material ending in a larger BH seed with masses up to $\sim$ few $10^5 M_\odot$ \cite{2006MNRAS.370..289B,2008MNRAS.387.1649B,2017ApJ...842L...6W}. Former simulations suggest that the conditions for the occurrence of this scenario, e.g., to reach metal-free gas able to form atomic (instead of molecular) gas, may be rare \cite{2016MNRAS.463..529H}. However, recent hydrodynamic N-body simulations show that DCBH scenarios are among the most preferred mechanisms to explain the origin of the SMBHs \cite{2022MNRAS.tmp.1507Z, Latif2022-wd}, albeit numerical resolution issues and the use of phenomenological recipes limit the generality of the results \cite{2022MNRAS.tmp.1507Z}.

On different physical grounds, channels of BH formation and growth associated with the early Universe (II) would take place before galaxy formation and include primordial BHs \cite{2020ARNPS..70..355C}, or even more exotic candidates such as topological defects in forms of cosmic string-loops \cite{2015JCAP...06..007B}. However, there is no observational evidence (nor direct or indirect) of such processes since they are associated with very early cosmological epochs poorly constrained by observations.

This work proposes an SMBH formation channel in the high $z$ Universe conceptually different from (I) and (II) discussed above. A crucial qualitative difference with the channels mentioned above is that it does not rely either on specific pristine gas assemblies or on the very early epoch of the Universe. Instead, it depends on the gravitational collapse and subsequent growth of dense fermionic DM cores that originate at the center of the halos as they form. Such novel \textit{dense core}-\textit{diluted halo} DM density distributions (profiles) are natural consequences of maximum entropy production principle (MEPP) scenarios of halo formation, in which the fermionic (quantum) nature of the particles is duly accounted for \cite{2021MNRAS.502.4227A, 2022IJMPD..3130002A}. 

This paper is organized as follows. In section \ref{sec:II}, we discuss the underlying physics behind our new SMBH seed formation scenario and give specific examples of how such seeds may arise in the high $z$ Universe. In section \ref{sec:III}, we present relativistic calculations of the time evolution of the mass and spin of the recently born BH seed with a surrounding accretion disk. We summarize and conclude in section \ref{sec:IV}.

\begin{figure*}
  \includegraphics[width=\hsize]{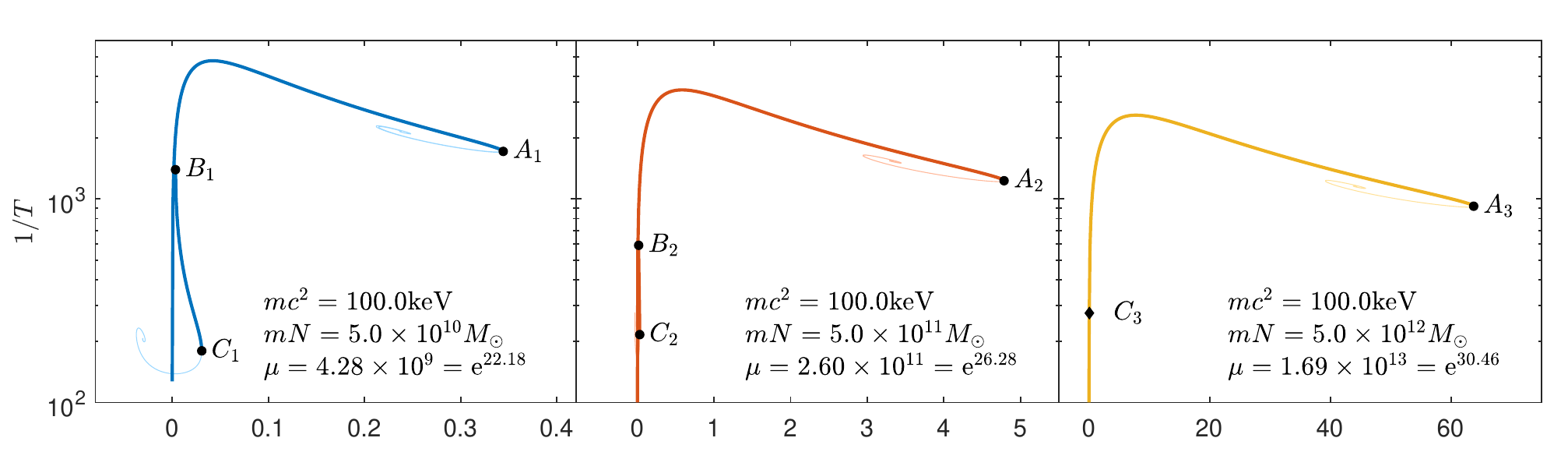}
  \includegraphics[width=\hsize]{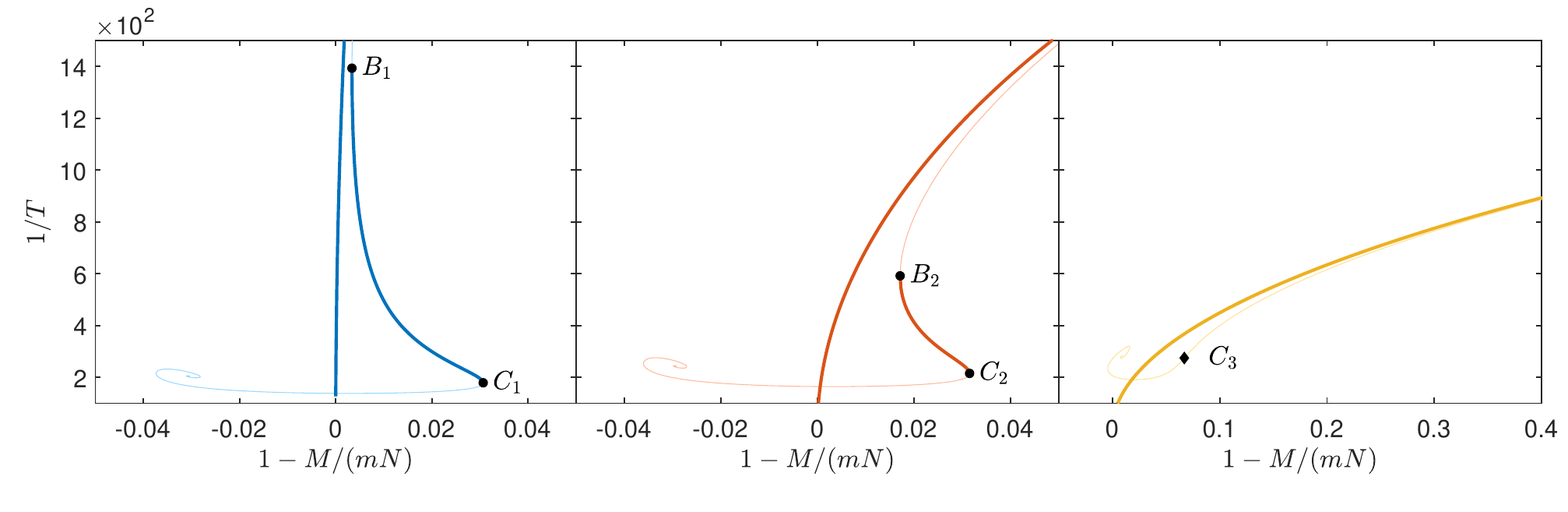}
  \caption{Caloric curves for the critical solutions of three examples with a darkino mass of $mc^2 = 100$ keV: (1) $M_{\rm vir} = 5\times10^{10} M_\odot$ and $r_{s} = 30$ kpc, (2) $M_{\rm vir} = 5\times10^{11} M_\odot$ and $r_{s} = 80$ kpc, (c) $M_{\rm vir} = 5\times10^{12} M_\odot$ and $r_{s} = 250$ kpc. The necessary constraints ($\mu = {\rm e}^{W_0 - \theta_0}$ and $N$) to compute the caloric curves are extracted from the corresponding critical solutions. For the x-axis, $1 - M/(mN)$ is (minus) the binding energy normalized with $mN$. For the y-axis, $T$ is the normalized temperature of the system as measured by an observer placed at infinity. Thin lines are unstable solutions, while thick lines are meta-stable. A black dot marks the change in stability. For example (3), there are no stable core-halo solutions. An equivalent solution with the same core mass as for the other examples ($C_1$ and $C_2$) is marked by the black diamond.}
  \label{fig:caloric-curve}
\end{figure*}

\begin{figure}
\center{\includegraphics[width=\hsize]{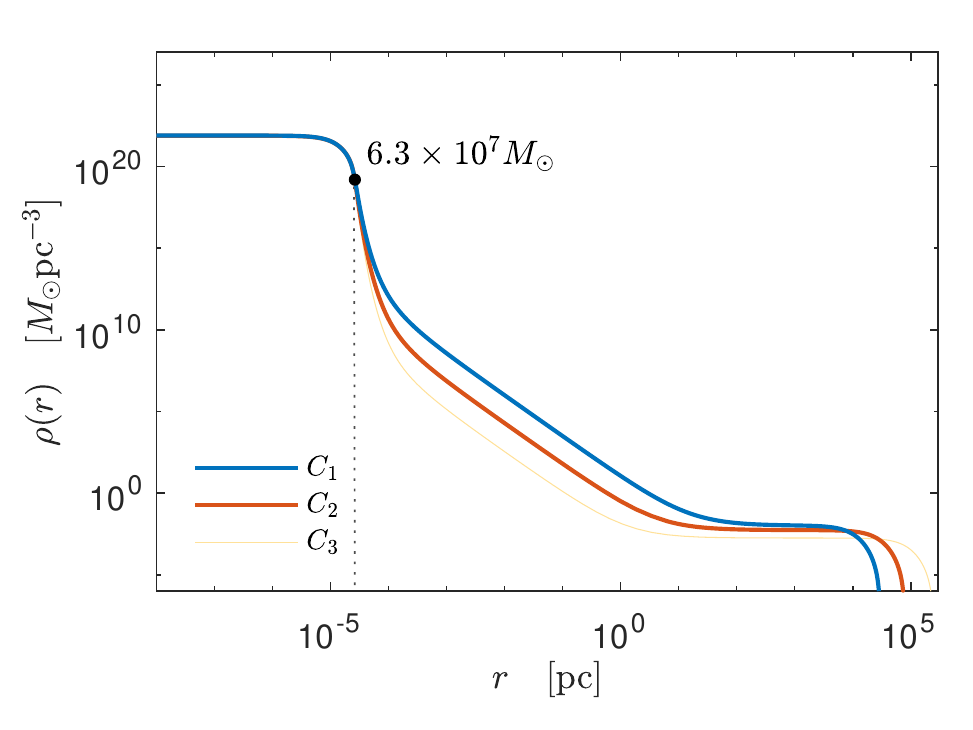}}
\caption{Density profiles of the solutions $C_1$, $C_2$ and $C_3$ from \cref{fig:caloric-curve}.  Thin lines are unstable solutions, while thick lines are meta-stable. The dotted line shows the corresponding fully degenerate core solution, providing the core radius $r_c$ where the density falls to zero (e.g., surface). The black point is labeled with the value of core mass $M_{\rm crit} = M(r_c)$.}
\label{fig:density-profiles}
\end{figure}

\begin{figure}
\center{\includegraphics[width=\hsize]{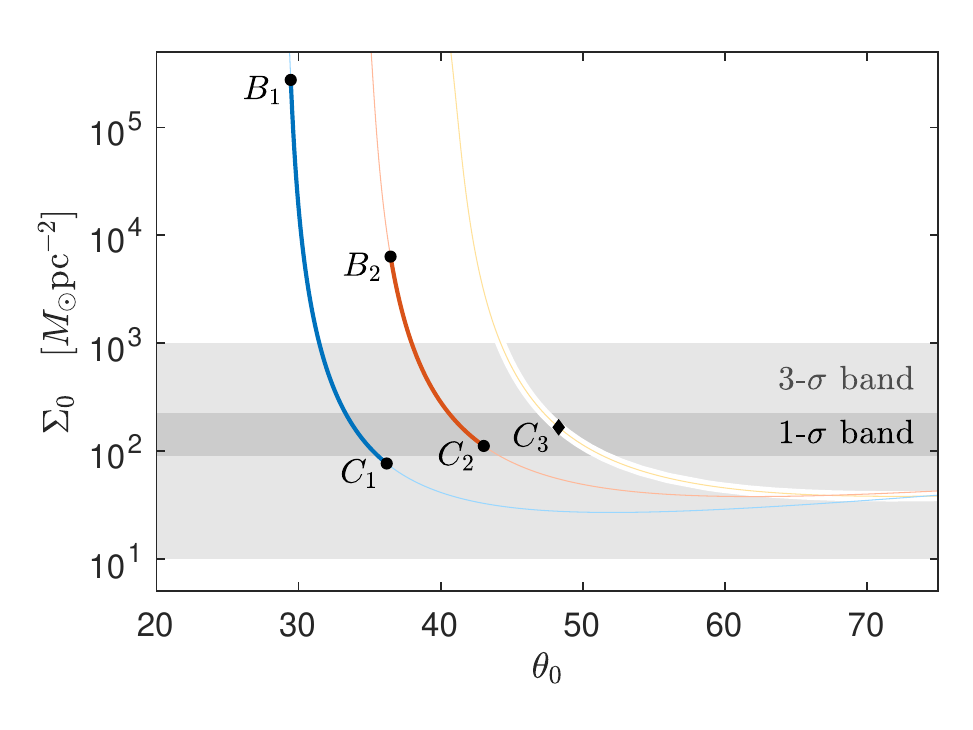}}
\caption{Surface density curves for three examples as described \cref{fig:caloric-curve}.  Thin lines are unstable solutions, while thick lines are meta-stable. The shaded regions mark the maximal extent of 1-$\sigma$ and 3-$\sigma$ errors as obtained in \citet{2009MNRAS.397.1169D}).}
\label{fig:surface-density-profiles}
\end{figure}
% \ref{sec:II}
%\ref{sec:III}

%%%%%%%%%%%%%%%%%%%%%%%%%%%%%%%%%%%%%%%%%%%%%%%%%%%%
%%%%%%%%%%%%%%%%%%%%%%%%%%%%%%%%%%%%%%%%%%%%%%%%%%%%
\section{The dark matter channel of SMBH formation}\label{sec:II}
%%%%%%%%%%%%%%%%%%%%%%%%%%%%%%%%%%%%%%%%%%%%%%%%%%%%
%%%%%%%%%%%%%%%%%%%%%%%%%%%%%%%%%%%%%%%%%%%%%%%%%%%%

The DM channel proposed for forming early BH seeds takes place within the more general and broad theory of DM halo formation as developed in \cite{2021MNRAS.502.4227A} within WDM cosmologies. Unlike the (zoom-in) hydrodynamical simulation approaches applied in baryonic BH-seed formation channels \cite{2022MNRAS.tmp.1507Z} (such as (Ia) and (Ib) introduced above), the problem of DM halo formation is here assessed using a thermodynamical approach for systems of self-gravitating fermions which maximize its coarse-grained entropy at the end of relaxation \cite{2021MNRAS.502.4227A}. This mechanism's most general density profile develops a dense and compact DM core (supported against gravity by Pauli degeneracy pressure) surrounded by a dilute halo. The first early studies of this kind of fermionic \textit{core}-\textit{halo} solutions dates back to the $'80$s \citep{1984ApJ...281..560C,1988NCimB.101..369I}, followed by a series of further and recent developments \cite{1990A&A...235....1G,1998MNRAS.296..569C,2002PrPNP..48..291B,2006IJMPB..20.3113C,2013NewA...22...39D,2014IJMPD..2342020A,2015MNRAS.451..622R,2015PhRvD..92l3527C,2018PDU....21...82A,2019PDU....24..278A,2020A&A...641A..34B,2021MNRAS.505L..64B,2022PhRvD.106d3538C,2022MNRAS.511L..35A,2023ApJ...945....1K}. The more realistic version of this model, which includes particle evaporation and central (fermion) degeneracy, was developed in General Relativity in \citet{2018PDU....21...82A} and is referred to as the (extended) Ruffini-Arg\"uelles-Rueda (RAR) model\footnote{It also called in the literature as the general relativistic fermionic King model \citep{2022PhRvD.106d3538C}.}. The model's fermionic halo explains the galaxy rotation curves, while the degenerate fermion core has key implications for galactic centers: it can mimic a central BH or eventually collapse into one \citep{2018PDU....21...82A,2019PDU....24..278A,2020A&A...641A..34B,2021MNRAS.502.4227A,2021MNRAS.505L..64B,2022MNRAS.511L..35A,2022IJMPD..3130002A}. 
One key advantage of this semi-analytical approach is that it allows for a detailed description of the relaxed halos from the very center to the periphery, not possible in N-body simulations (due to limitations in resolution at inner-halo scales). Moreover, the thermodynamic approach here applied includes richer physical ingredients than those involved in traditional simulations: (i) general relativity (GR) --- necessary for a proper gravitational DM core-collapse towards an SMBH seed; (ii) the quantum nature of the particles --- allowing for an explicit fermion mass dependence in the profiles; (iii) the Pauli principle self-consistently included in the phase-space DF at relaxation --- giving place to novel \textit{core-halo} profiles. Interestingly, this theoretical framework allows linking the behavior and evolution of the dark matter particles from the early Universe to the late stages of non-linear structure formation at virialization. That is, the DM halo profiles are obtained by first calculating the linear matter power spectrum for $\mathcal{O}(\text{keV})$ DM fermions, to then use the corresponding extended Press-Schechter formalism to obtain the virial halo mass, $M_{\rm vir}$, with associated redshift $z_{\rm vir}$ (see Appendix, and \cite{2021MNRAS.502.4227A} for a previous result). Finally, we obtain the fermionic halos by assuming that a MEPP takes place at the end of relaxation and agrees with the virial mass constraints. Such a MEPP, originally introduced in \cite{1998MNRAS.300..981C} generalizing Lynden-Bell results, allows obtaining a most likely coarse-grained DF of Fermi-Dirac type\footnote{Approaches of this kind based on statistical mechanics may present potential difficulties: the relatively short time-scale involved during violent relaxation may not be enough for the system to explore the full phase-space to reach a most likely final state (see \citealp{2022PhRvD.106d3538C}).} [see Eq. (1) in \citet{2021MNRAS.502.4227A}] that depends on four free parameters, $m$ the particle mass, $\beta$ the (dimensionless) temperature, $\theta$ the degeneracy parameter, and $W$ the cut-off particle energy. All parameters are set at the center of the configuration (denoted with the subscript $0$) to fully solve the system of equilibrium differential equations of the RAR model [see Eqs. (8)--(12) in \citet{2021MNRAS.502.4227A}]. The RAR model can be applied to form a DM halo and their central SMBH seeds.

Once having such a DF at the end of the relaxation, we calculate the full family of fermionic density profiles at equilibrium in GR, all with given total particle number N and thus with the same total (Newtonian) halo mass ($M_{\rm tot}\equiv M_{\rm vir}=m N$). For such an endeavor, we follow the thermodynamic approach applied in \cite{2021MNRAS.502.4227A} within the microcanonical ensemble to further calculate the important problem of (thermodynamical and dynamical) stability of such a family of equilibrium solutions. That is, not all the equilibrium solutions of self-gravitating fermions will be thermodynamically stable\footnote{While the problem of equilibrium involves the extremization of entropy $\delta S=0$ (at fixed energy and N), the problem of stability has to do with second-order variations of entropy $\delta^2 S=0$ \cite{2020EPJP..135..290C}.}. This is done following the \textit{Katz criterium} (see Appendix A in \cite{2021MNRAS.502.4227A} and Appendix C in \cite{2020EPJB...93..208A} for updated summaries, and \cite{katz1978number,katz1979number} for the original works) for which is necessary to calculate the caloric curves in GR, given as the inverse temperature of the system $1/\hat T\equiv \beta^{-1}$ Vs. (minus) the binding energy $-E_b=-(M-mN)c^2$ (with $\hat T$ the normalized temperature such that $\beta=k T/(m c^2)$ and $k$ the Boltzmann constant). A distinctive characteristic of such general relativistic caloric curves for fermions (at difference with the Newtonian case), is the existence of a \textit{last stable configuration} located at the turning point (see points $C_i$ in \cref{fig:caloric-curve}), followed by a second spiral feature of relativistic origin. Based on the Katz criterium, this important result was first shown in \citet{CHAVANIS2020135155,2020EPJP..135..290C,2020EPJB...93..208A,2022PhRvD.106d3538C} for a self-gravitating system of fermions bounded in a box, together with a detailed characterization of the caloric curves and the gravitational phase transitions occurring to the Fermi gas. Remarkably, as shown as well in \cite{2021MNRAS.502.4227A} for the more realistic RAR (or relativistic fermionic King) model, the existence of such a \textit{last stable configuration} located at point $C_i$, implies the onset of a thermodynamical instability of the \textit{core-halo} solutions, where the fermion-core collapses towards an SMBH. 

%which is still surrounded by the halo. 

% textbf{as first demonstrated in [Chavanis papers] within GR for box-confined configurations, and extended here () under the more realistic RAR model in \cite{2021MNRAS.502.4227A}}

Based on the above (relativistic) thermodynamical analysis, we demonstrate, for the first time, the existence of a critical fermion-core (located at $C_i$ in the caloric curve) which is surrounded by a DM halo of realistic astrophysical application (see the DM density profiles in \cref{fig:density-profiles}, concerning \cref{fig:caloric-curve} and the fulfillment with observations in \cref{fig:surface-density-profiles}). We do it for a typical particle mass in the range $50$--$345$ keV, i.e., $m=100$ keV, to then explore (see \cref{subsec:IIIB}) other particle masses in that range. The relevance of such a narrow window of particle masses is taken from \cite{2018PDU....21...82A,2019PDU....24..278A}, where it was shown it is possible to find core-halo RAR solutions where the outer halo agrees with the galaxy rotation curves while the DM core (not necessarily critical) can mimic the central BH (see also \cite{2020A&A...641A..34B,2021MNRAS.505L..64B,2022MNRAS.511L..35A} for a tailored analysis about the Milky Way and Sgr A*). 

In \cref{fig:caloric-curve}, we give three specific examples of caloric curves for different (Newtonian) halo masses $M_{\rm vir}$ covering the relevant range between $5\times10^{10}$--$5\times10^{12} M_\odot$. Among all the equilibrium \textit{core-halo} solutions along each caloric curve, only the ones placed within the branches $B_i - C_i$ ($i=1,2,3$) are thermodynamically and dynamically stable within cosmological time-scales, as clearly explained in \cite{2021MNRAS.502.4227A} (see also \citep{CHAVANIS2020135155,2020EPJP..135..290C,2020EPJB...93..208A,2022PhRvD.106d3538C} for analogous results obtained for fermionic systems bounded in a box). We recall that the \textit{core-halo} solutions located at $C_i$ (see \cref{fig:density-profiles}) correspond to the last stable configuration where the DM-core achieves the onset of gravitational collapse (of relativistic origin) towards a BH. Interestingly, at the fixed mass of $m=100$ keV, the theory predicts a threshold total halo mass $M_{\rm vir}\sim 5\times 10^{12} M_\odot$ above which the stable branch $B_3 - C_3$ disappeared (see right panels of \cref{fig:caloric-curve}). The shrinking of the metastable branch ($B-C$) as the total mass $M_{\rm tot}$ (or $N$) increases; together with the existence of a threshold particle number $N^{*}$ above which the meta-stable branch disappear, was first shown in \cite{CHAVANIS2020135155,2020EPJB...93..208A} for box confined systems. This remarkable result, when applied to realistic halos as in this work, may explain why we do not observe single virialized galaxies above such an order of magnitude (i.e., above $\mathcal{O}(10^{12}) M_\odot$), indicating how powerful the thermodynamics of self-gravitating systems can be.
% \textbf{We refer to \cite{CHAVANIS2020135155,2020EPJB...93..208A} for first results obtained in systems bounded in a box}

Moreover, in such a halo-mass window, we further show in \cref{fig:surface-density-profiles} that the halo regime of the RAR solutions has the required morphology in the sense of being able to fulfill the DM surface density relation. We also show in \cref{fig:density-profiles} the density profiles of such core-halo astrophysical solutions at the onset of DM core-instability, all having a typical SMBH seed of $M_{\rm crit} \approx 6.3 \times 10^7 M_\odot$. We have defined the SMBH seed mass at the core radius $r_c$ of the  core-halo solution, i.e., $M_{\rm crit} = M(r_c)$, with $r_c$ coinciding with the surface radius of the corresponding fully degenerate solution (e.g., where the density falls to zero, see dotted line in \cref{fig:density-profiles}). Such a numerical value of the critical mass can be well approximated with the semi-analytic \cref{eq:Mcrit} (only valid within the fully degenerate regime), which is no other than the Oppenheimer-Volkoff (OV) mass limit \cite{1939PhRv...55..374O}
\begin{equation}\label{eq:Mcrit}
M_{\rm crit} \approx 0.384 \frac{m_{\rm Pl}^3}{m^2} \approx 6.274 \times 10^9 \left( \frac{\text{10 keV}}{m c^2}  \right)^2 M_{\odot},
\end{equation}
where $m_{\rm Pl} = \sqrt{\hbar c/G} \approx 2.176 \times 10^{-5}$ g is the Planck's mass and $m$ is the darkino mass.

The reason for the validity of this critical mass approximation of our DM cores can be easily understood when realizing that the core-halo fermionic solutions under consideration here (see \cref{fig:density-profiles}) encompass two different regimes: a highly degenerate (quantum) regime of the fermionic-core (i.e., $\theta_0>10$) close to the fully-degenerate case, which monotonically transitions to the classical regime at larger distances from the center, leading to the (Boltzmannian) halo region (where $\theta(r)\ll -1$). Further detailed explanations about the equivalence between the traditional turning-point instability criterium of core-collapse \cite{2014CQGra..31c5024S}, to that of the last (dynamical and thermodynamical) stable solution at point $C_i$ in the caloric curves are given in Section 4 of \cite{2021MNRAS.502.4227A} and references therein.

%%%%%%%%%%%%%%%%%%%%%%%%%%%%%%%%%%%%%%%%%%%%%%%%%%%%
%%%%%%%%%%%%%%%%%%%%%%%%%%%%%%%%%%%%%%%%%%%%%%%%%%%%
\section{BH mass and spin evolution}\label{sec:III}
%%%%%%%%%%%%%%%%%%%%%%%%%%%%%%%%%%%%%%%%%%%%%%%%%%%%
%%%%%%%%%%%%%%%%%%%%%%%%%%%%%%%%%%%%%%%%%%%%%%%%%%%%

We follow the treatment of a nearly geodesic thin accretion disk around a Kerr BH in \cite{1973blho.conf..343N, 1974ApJ...191..499P, 1974ApJ...191..507T}. Matter and radiation transfer energy and angular momentum to the BH during the accretion. In particular, it is essential to account for the feedback of radiation/photons onto the BH since they exert a counter-torque \cite{1970PhRvD...1.2721G} that avoids the BH from reaching the extreme regime $a=M$. This implies that the accretion of massive particles and radiation does not lead the BH to become a naked singularity \cite{1974ApJ...191..507T}. We denote by $dm$ the rest-mass accreted by the BH in a coordinate time interval $dt$, so $\dot{m} = dm/dt$ is the rest-mass accretion rate, and $\dot{M}_{\rm rad}$ and $\dot{J}_{\rm rad}$ are the rate of energy and angular momentum transfer by radiation to the BH. We refer the reader to the appendix \ref{app:B} for details of the equations governing the evolution of the mass and angular momentum of the BH. We use geometric units $c=G=1$ unless otherwise specified.

%%%%%%%%%%%%%%%%%%%%%%%%%%%%%%%%%%%%%%%%%%%%%%%%%%%%%
\subsection{Accretion rate and luminosity}\label{sec:IIIA}
%%%%%%%%%%%%%%%%%%%%%%%%%%%%%%%%%%%%%%%%%%%%%%%%%%%%%

We calculate the rate at which rest-mass flows inward through the local balance between the tidal gravitational acceleration and the radiation pressure along the $z$ coordinate. This condition is
\begin{equation}\label{eq:localcond}
zR^{\tilde{z}}_{\tilde{0}\tilde{z}\tilde{0}} = \kappa F(r),
\end{equation}
where $\kappa = 0.34$ cm$^2$ g$^{-1}$ is the Thomson electron scattering opacity, $\boldsymbol{R}$ is the Riemann tensor. Using the change of variable $x=\sqrt{r/M}$, Eq.~\eqref{eq:localcond} becomes
\begin{equation}\label{eq:localcond2}
\frac{3\,\kappa\,\dot{m}}{8\pi z}f(x,\alpha)= 1,
\end{equation}
where
\begin{equation}\label{eq:fexp}
    f(x,\alpha) = \frac{x^{3}\, g(x,\alpha)}{x^4-4\alpha x+3\alpha^2},
\end{equation}
and
\begin{equation}\label{eq:gexp}
    g(x,\alpha) = x - x_0 - \frac{3}{2}\alpha\ln\left(\frac{x}{x_0}\right) - 3(A_1+A_2+A_3),
\end{equation}
with
\begin{equation}\label{eq:Aexp}
    A_1 = \frac{(x_1-\alpha)^2}{x_1(x_1-x_2)(x_1-x_3)}\ln\left(\frac{x-x_1}{x_0-x_1}\right).
\end{equation}
Here, $x_{1},x_{2},x_{3}$ are the roots of the polynomial $x^3-3x+2a$. The terms $A_2$ and $A_3$ can be obtained from $A_1$ through a cyclic order of the set $\{1,2,3\}$. For a given value of $\alpha$ and an appropriate expression $z=z(r)$, the value of $\dot{m}$ such that Eq.~\eqref{eq:localcond2} has exactly one solution defines a critical accretion rate $\dot{m}_{\rm crit}$. Above this rate, the disk enters the super-Eddington regime, and the thin disk approximation breaks down. The tidal gravitational pull reaches a maximum value at $z=H$. Additionally, the thin disk approximation is reasonably accurate insofar as the half thickness of the disk obeys $H < r$~\cite{Chen_2007,Liu:2017kga}. Consequently, by introducing a parameter $0<\beta \leq 1$ and setting $z=r$~(see \cite{2015MNRAS.454.3432A} for details), we calculate the accretion rate by 
\begin{equation}
    \dot{m} \equiv \beta\,\dot{m}_{\rm crit} = \frac{8\pi\beta M}{3\kappa\,{\rm max}\left\{\frac{f(x,\alpha)}{x^2}\right\}},\label{eq:accrate}
\end{equation}
with ${\rm max}\left\{\frac{f(x,\alpha)}{x^2}\right\}$, the maximum value can take the given function at each radius for given $\alpha$. To calculate the power emitted by the system, we only consider the photons that leave the disk and do not fall into the BH. The procedure is equivalent to calculating the rate of energy transfer to the BH, i.e., Eq.~\eqref{eq:Mdotmatter}, but using the factor $1-C$ instead of $C$
\begin{equation}\label{eq:luminosity}
L_{\rm source}=-\frac{2}{\pi}\int_{r_{0}}^{\infty}\!\int_{0}^{\pi/2}\!\!\int_ {0}^{2\pi}\!\!\left(1-C\right)k_{t}\!F(r)dS.
\end{equation}

We do not consider the possible photon recapture by the disk.

%%%%%%%%%%%%%%%%%%%%%%%%%%%%%%%%%%%%%%%%%%%%%%%%%%%%%
\subsection{Growth of Kerr BH seeds}\label{subsec:IIIB}
%%%%%%%%%%%%%%%%%%%%%%%%%%%%%%%%%%%%%%%%%%%%%%%%%%%%%

\begin{figure*}%[!hbtp]
\centering
\includegraphics[width=0.49\hsize,clip]{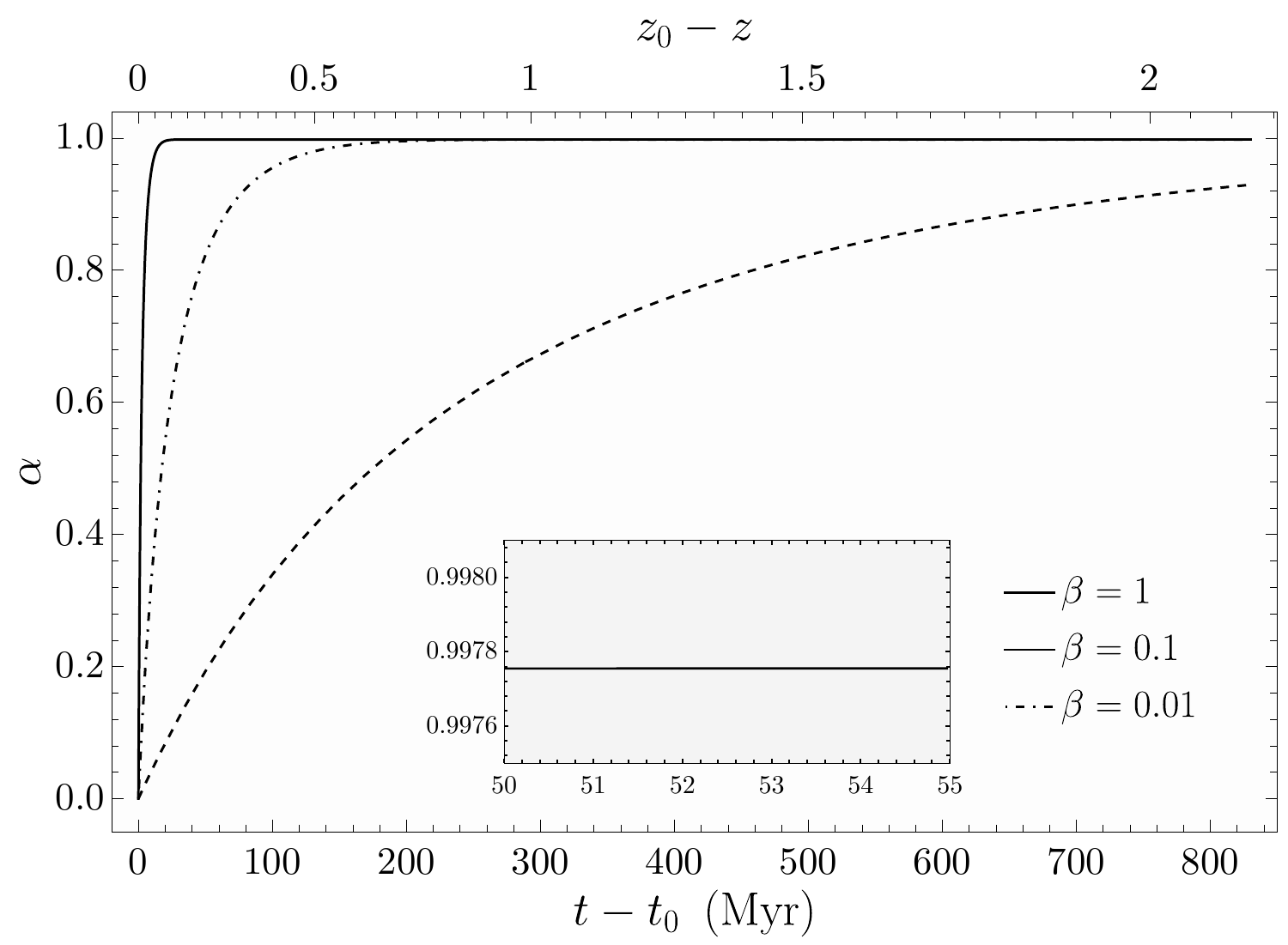}\includegraphics[width=0.5\hsize,clip]{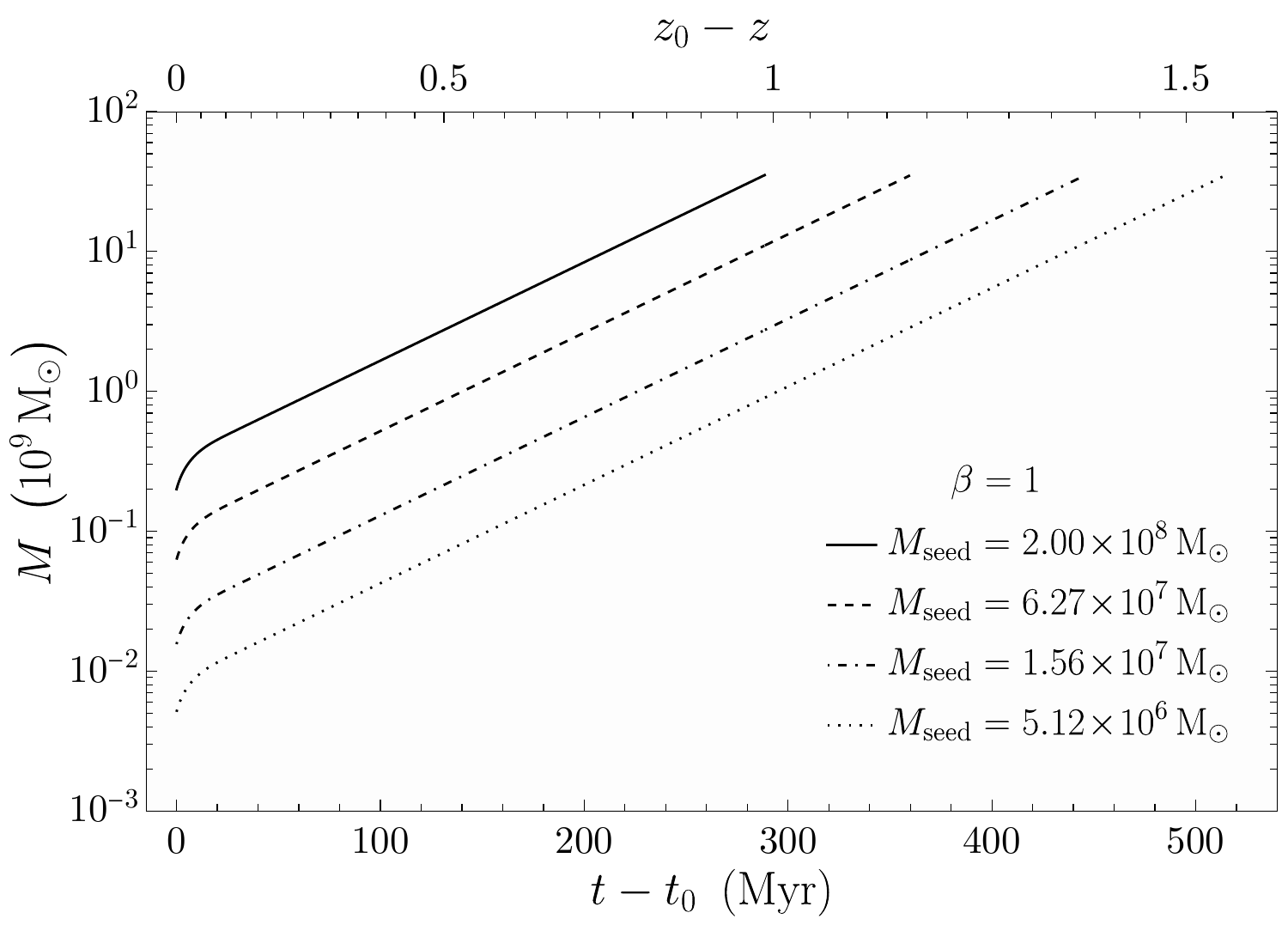}\\
\includegraphics[width=0.5\hsize,clip]{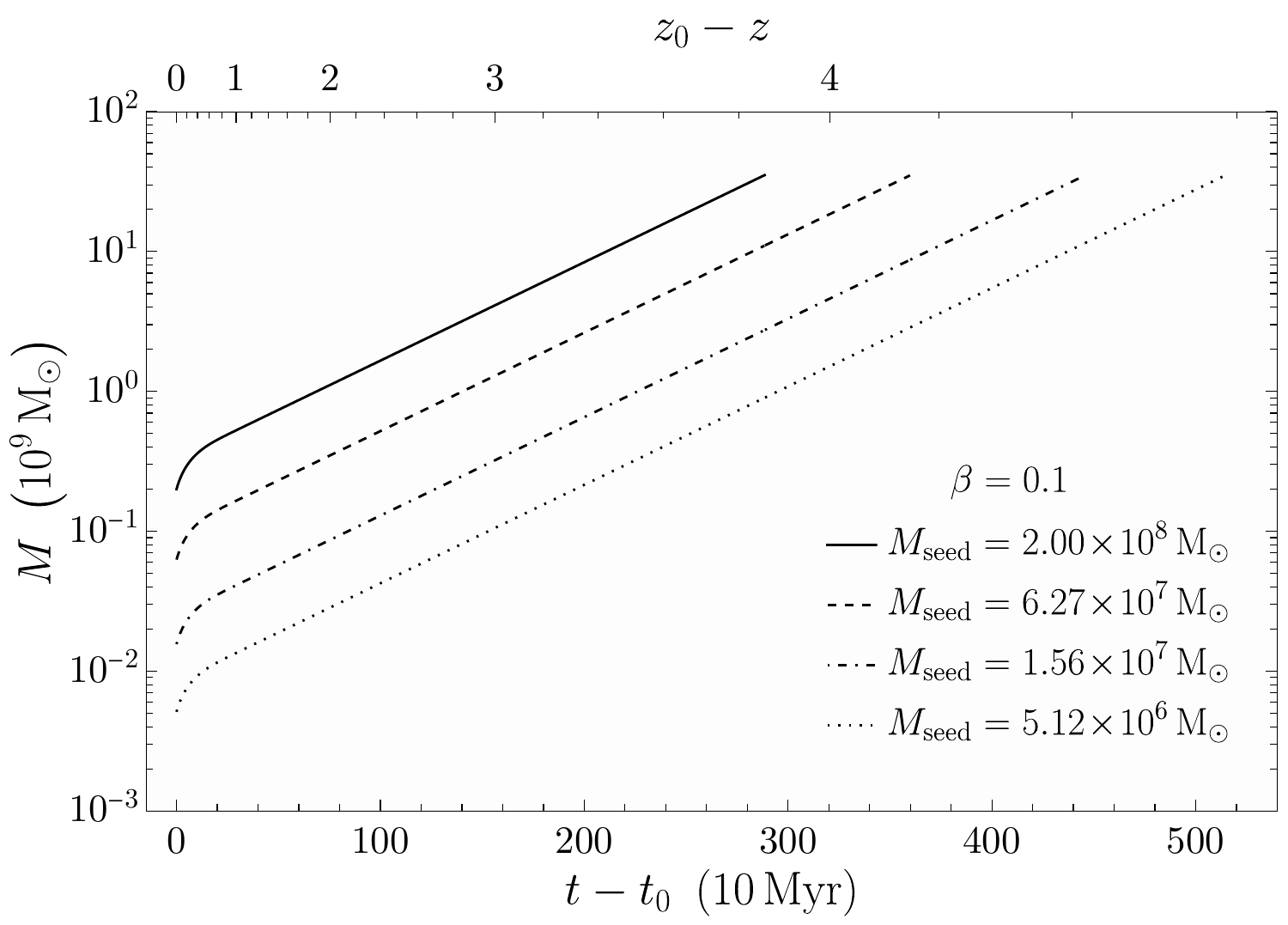}\includegraphics[width=0.5\hsize,clip]{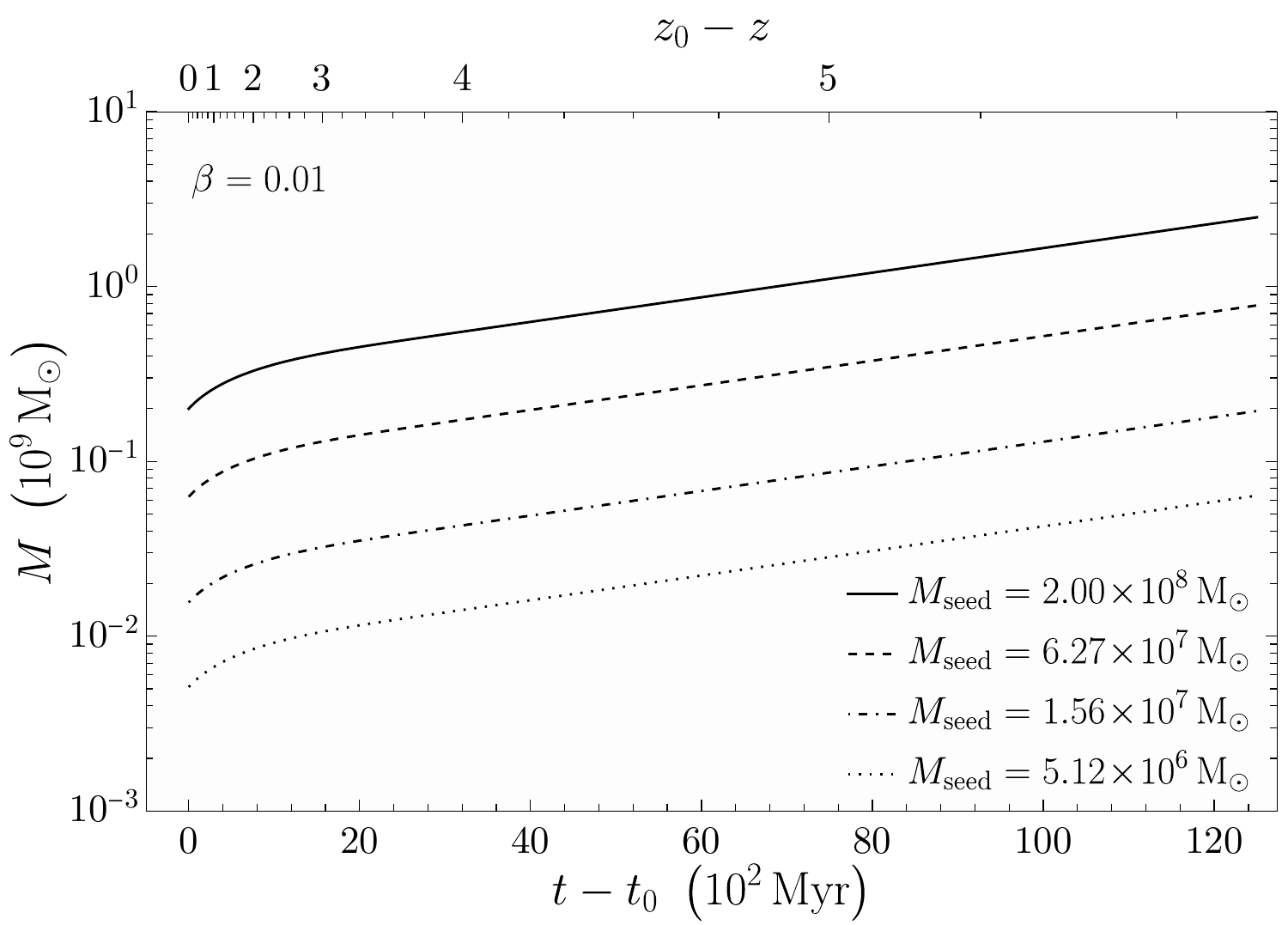}
\caption{BH evolution in time for different BH seeds and values of $\beta$. Initial conditions are $\alpha_i=0$ and darkino masses $56$~keV, $100$~keV, $200$~keV, and $350$~keV. The spin parameter does not depend on the BH mass. The initial redshift is $z_0=5.5$ with $t_0=1022$ Myr for a halo mass $M_{vir}=5\times 10^{11}M_{\odot}$.}
\label{fig:bh_evol}
\end{figure*}

\begin{figure*}%[!hbtp]
\centering
\includegraphics[width=0.5\hsize,clip]{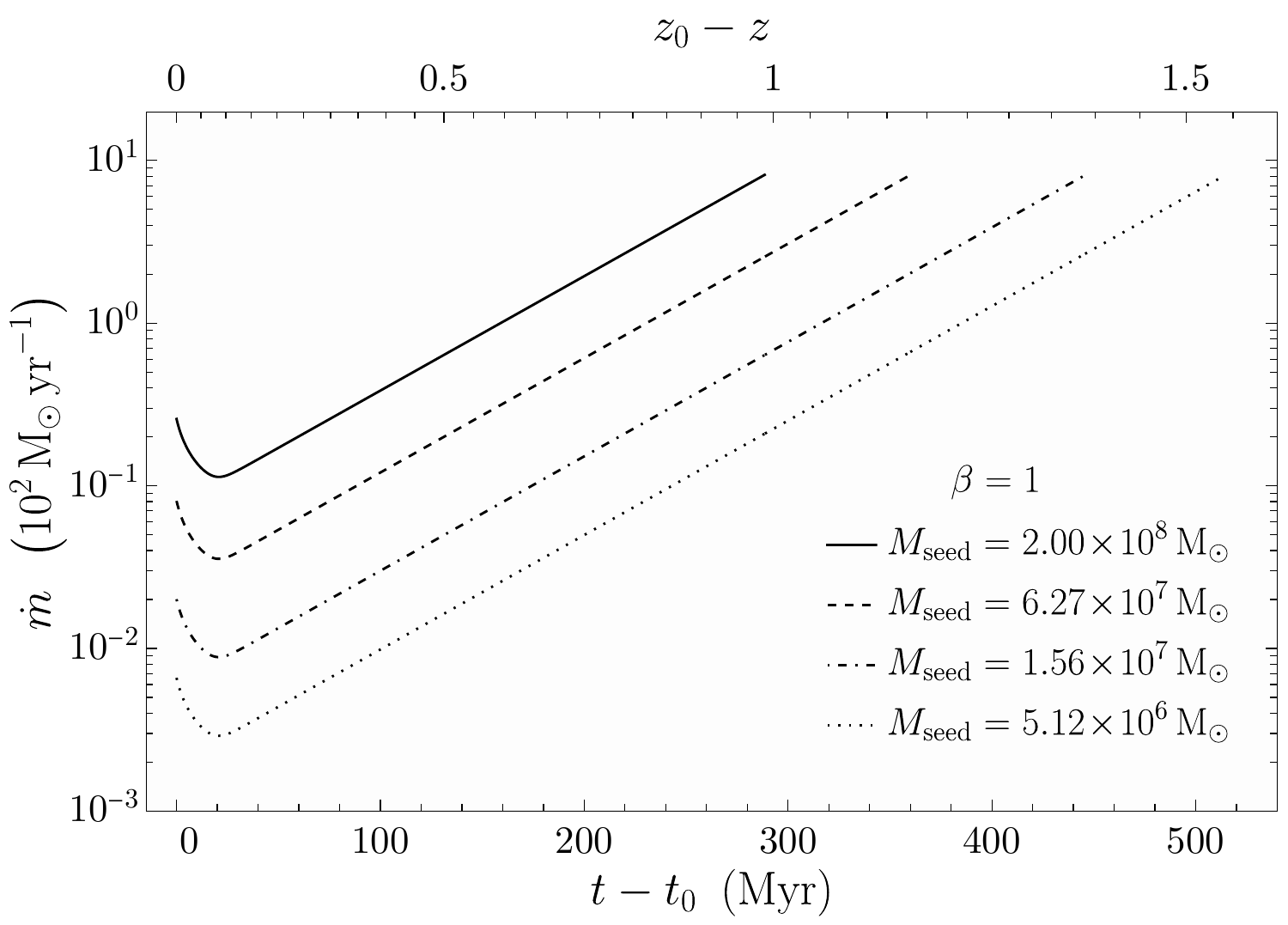}\includegraphics[width=0.5\hsize,clip]{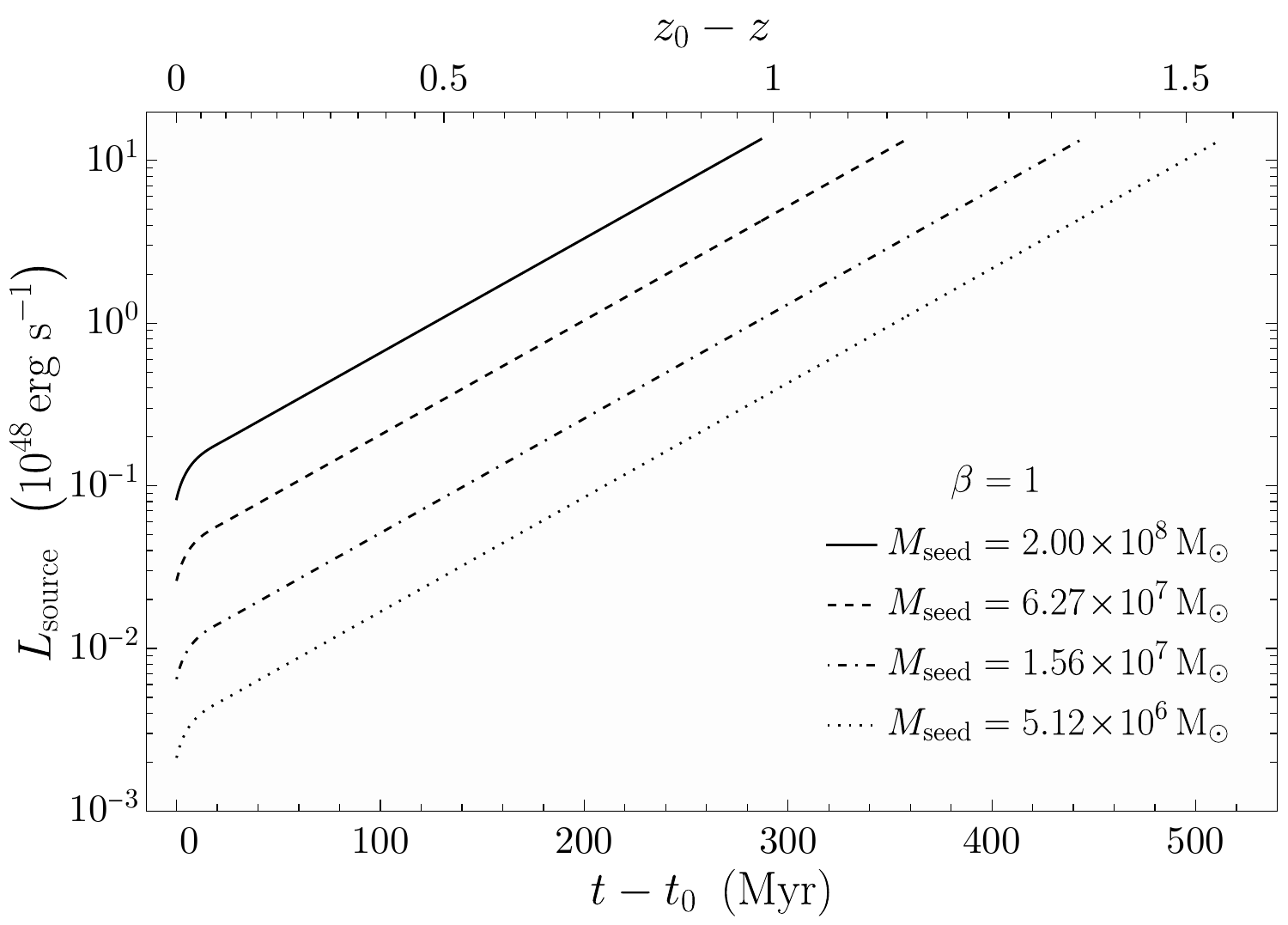}
\caption{Accretion rate and Luminosity at source as functions of $t$ for different BH seeds with $\beta=1$. Initial conditions are $\alpha_i=0$ and darkino masses $56$~keV, $100$~keV, $200$~keV, and $350$~keV. The initial redshift is $z_0=5.5$ with $t_0=1022$ Myr for a halo mass $M_{vir}=5\times 10^{11}M_{\odot}$.}
\label{fig:accr_lum_avol}
\end{figure*}

The equations of evolution are
\begin{align}
    \dot{M} &= \dot{m}\epsilon_0 + \dot{M}_{\rm rad},\label{eq:mass_evol}\\
    \dot{\alpha} &= \frac{1}{M^2}\left(l_0\dot{m} + \dot{J}_{\rm rad} -2M\dot{M}\alpha\right).\label{eq:alpha_evol}
\end{align}

It is helpful to make the $M$ dependence explicit and express the equations in terms of the system's parameters and dimensionless variables. To that end, let us write
\begin{align}
    M &= M_{i}\tilde{M}, \label{eq:adbhmass}\\
    \dot{m} &= M_{i}\Tilde{M}\beta \dot{\tilde{m}},\label{eq:adaccrate}\\
    \dot{M}_{\rm rad} &= M_{i}\Tilde{M}\beta \dot{\tilde{m}}\tilde{M}_{\rm rad},\label{eq:adMdotrad}\\
    \dot{J}_{\rm rad} &= M_{i}^{2}\Tilde{M}^{2}\beta \dot{\tilde{m}}\tilde{J}_{\rm rad},\label{eq:adJdotrad}\\
    l_{0} &= M_{i}\Tilde{M}\tilde{l}_{0},\label{eq:adangmom}
\end{align}
where $\dot{\tilde{m}}$, $\tilde{M}_{\rm rad}$, $\tilde{J}_{\rm rad}$ and $\tilde{l}_{0}$ are functions of $\alpha$ only. With these definitions, the ratio between Eq.~\eqref{eq:mass_evol} and Eq.~\eqref{eq:alpha_evol} leads to the solution
\begin{equation}
    \tilde{M} = \exp\left\{\!\int_{\alpha_i}^{\alpha}\!\frac{ \epsilon_0 + \tilde{M}_{\rm rad}}{\tilde{l}_0 +\tilde{J}_{\rm rad}-2\alpha_{*}\!\left(\epsilon_0 + \tilde{M}_{\rm rad}\right)}d\alpha_{*}\!\!\right\}\!,\label{eq:mass_sol}
\end{equation}
for the BH mass as a function of the dimensionless spin parameter. From Eq.~\eqref{eq:alpha_evol} we obtain
\begin{equation}
     \Delta t = \frac{1}{\beta}\!\int_{\alpha_i}^{\alpha}\frac{1}{\dot{\tilde{m}}\left[\tilde{l}_0+\tilde{J}_{\rm rad} - 2\alpha_{*}\left(\epsilon_0+\tilde{M}_{\rm rad}\right)\right]}d\alpha_{*}.\label{eq:alpha_sol}
\end{equation}

From Eq.~\eqref{eq:mass_sol} and Eq.~\eqref{eq:alpha_sol}, we can deduce some properties of the BH's temporal evolution: The solution $\alpha(t)$ is independent of $M$, but it does depend on the parameter $\beta$. Moreover, the general solution $\alpha(t,\beta)$ obeys the relation $\alpha(t,\beta) = \alpha(\beta t,1)$. The BH mass inherits this property through Eq.~\eqref{eq:mass_sol}, and we get $M(t,\beta) = M(\beta t,1)$. The same will happen with any $\alpha$ and $M$ function. This behavior suggests that knowing the solution for $\beta=1$ is enough to produce other solutions by properly scaling the variable $t$ and multiplying it by an adequate power of $\beta$. For example, the accretion rate and the power emitted obey the relations $\dot{m}(t,\beta)=\beta\dot{m}(\beta t,1)$ and $L_{\rm source}(t,\beta)=\beta L_{\rm source}(\beta t,1)$.

The properties of $M(t,\beta)$ and $\alpha(t,\beta)$ become evident when inspecting the panels in \cref{fig:bh_evol}, where we show the evolution of several BH seeds for values $\beta = 1$, $\beta = 0.1$, and $\beta = 0.01$ as a function of time and the cosmological redshift. The initial redshift $z_0$ corresponds to the typical collapse $z^*$ of a gravitationally bound object with mass $M_{\rm vir}$ in the extended Press-Schechter formalism (see \cref{sec:AppA}). The standard $\Lambda$CDM cosmology sets the relation between $t$ and $z$. Note that $\alpha(t,\beta)$ does not depend on the BH mass, so the top left panel contains only three curves, one for each value of $\beta$.

In \cref{fig:accr_lum_avol}, we show the evolution of the accretion rate and the power emitted by the BH. The slight reduction in the accretion rate for small values of $t$ occurs because the factor ${\rm max}\left\{f(x,\alpha)/x^2\right\}$ in Eq.~\eqref{eq:accrate} increases by a factor of $\sim 5$ as $\alpha$ grows. The spin-up of a BH from $\alpha_i = 0$ to $\alpha = 0.99775$ takes $\sim 37/\beta$ Myr. The BH mass grows from $M_{i}$ to $\sim 3M_{i}$. Once $\alpha$ reaches the state $\dot{\alpha} \approx 0$, the BH mass grows exponentially. By solving Eq.~\eqref{eq:mass_evol}, we find a relation that allows us to estimate the time $\Delta t$ needed for any BH seed to grow up to a final mass $M_f > 3M_i$, starting from $\alpha_i = 0$:
\begin{equation}
\Delta t= 6.2\times10^{7}\ln\left\{\frac{3M_{f}}{5M_{i}}\right\}\beta^{-1}\; \text{yr}.
\label{eq:time_approx}
\end{equation}

Our result differs from Eq.~(1) in \citet{Haiman_2001}, viz.,
\begin{equation}
\Delta t_{\rm HL} = 4\times 10^{8}\varepsilon\ln\left\{\frac{M_{f}}{M_{i}}\right\}\beta^{-1} \text{ yr},
\label{eq:haiman_loeb}
\end{equation}
where $\varepsilon = L_{\rm source}/\dot{m}c^2$ is the radiative efficiency of the accretion process. The differences arise for two reasons: first, \citet{Haiman_2001} did not consider the BH spin, while our treatment accounts for it in a self-consistent manner: \cref{fig:bh_evol} shows that during the transition to a saturated spin, the BH growth is faster than exponential. Second, their definition of the accretion rate rests upon a \textit{global} balance between Newtonian gravity and spherically symmetric radiation pressure. Consequently, the radiative efficiency appears as a constant factor in \cref{eq:haiman_loeb}. Our definition adopts a relativistic \textit{local} balance between radiation pressure and vertical gravity, which increases the accretion rate~\cite{2015MNRAS.454.3432A} and allows the evolution of the efficiency with the BH parameters $M$ and $\alpha$. 

Using the limiting efficiency for a Kerr BH, $\varepsilon = 0.3$, we obtain $\Delta t < 0.52\Delta t_{\rm HL}$. Thus, the growth from a possible BH seed within our framework of $M_i=5\times 10^{6}M_{\odot}$ to $M_f=5\times 10^{9}M_{\odot}$ takes only $\Delta t \approx 0.4$~Gyr. 
%($\Delta t_{\rm HL} \approx 1020$~Myr).
In particular, our DM channel for SMBH formation predicts that typical DM halos of $M_{\rm vir}\sim 10^{11} M_\odot$ formed at $z_0\sim 7.5$ (see \cref{sec:AppA}) can harbor SMBH seeds of $M_i= 6.3\times 10^7 M_\odot$ (i.e., for $m=100$ keV), which can grow (within standard accretion rates) up to $M_f \equiv M=3\times 10^9 M_\odot$ in $\Delta t\approx 0.2$ Gyr, thus in agreement with most distant (i.e., $z \sim 6$) and most massive quasars observed (see, e.g., \cite{2022NewAR..9401642M}). This result provides a new channel for SMBH formation from DM which can overcome traditional baryonic scenarios such as Pop. III stars, whose light BH seeds of $< 10^3 M_\odot$ fail to grow even to $\sim 10^8 M_\odot$ by $z\sim 6$ \citep{2022MNRAS.tmp.1507Z}.

%%%%%%%%%%%%%%%%%%%%%%%%%%%%%%%%%%%%%%%%%%%%%%%%%%%%
\section{Conclusions}\label{sec:IV}
%%%%%%%%%%%%%%%%%%%%%%%%%%%%%%%%%%%%%%%%%%%%%%%%%%%%
%%%%%%%%%%%%%%%%%%%%%%%%%%%%%%%%%%%%%%%%%%%%%%%%%%%%

We have proposed a novel channel for SMBH formation in the high redshift Universe, which is not associated with baryonic matter (massive stars) or primordial cosmology. Instead, it relies on the gravitational collapse into a BH of fermionic dense DM cores that arise at the center of DM halos as they form and on the subsequent growth of the newborn BH by accretion. The formation of dense core-dilute halo density distributions of DM form when accounting for a fermionic (quantum) nature of the DM particles in the structure of the DM halos, which is not feasible in traditional N-body simulations \cite{2021MNRAS.502.4227A, 2022IJMPD..3130002A}. For fermion masses in the range of $50$--$345$ keV, this alternative non-linear structure formation approach predicts stable DM halos that agree with observations and harbor dense DM cores at the brink of gravitational collapse, with masses on the range of $10^6$--$10^8 M_\odot$ (see Section \ref{sec:II} for a WDM cosmology with $m=100$ keV). Thus, it offers a whole new range of SMBH seeds that are considerably larger than the ones predicted by baryonic formation channels, including the DCBH scenario (see Section \ref{sec:I}).

In this article, we assessed the mass and angular momentum evolution of such massive BH seeds using a standard, geodesic general relativistic disk accretion model. We self-consistently account for the feedback of radiation/photons onto the BH (see Section \ref{sec:III}). We have explicitly shown in Section \ref{sec:IV} that these SMBH can grow to masses in the range $\sim 10^9$--$10^{10} M_\odot$ within the first Gyr of the life of the Universe, in good agreement with the farthest quasars observed, without invoking unrealistic (or fine-tuned) accretion rates. A relevant advantage of the present framework is that it does not require star formation within such a short cosmological time scale as in traditional baryonic channels. At the same time, it naturally connects the total mass of a host galaxy and the mass of its central SMBH observed today, all in terms of DM and its cosmological evolution.   

 \section*{Acknowledgements}

C.R.A. was supported by CONICET of Argentina, the ANPCyT (grant PICT-2018-03743), and ICRANet. K.B. acknowledges partial support from the Science Committee of the Ministry of Science and Higher Education of the Republic of Kazakhstan (Grant No. AP19680128).

\section*{DATA AVAILABILITY}
The article's data will be shared with the corresponding author upon reasonable request.

\bibliographystyle{mnras}
\bibliography{references}
%%%%%%%%%%%%%%%%%%%%%%%%%%%%%%%%%%%%%%%%%%%%%%%%%%%%
\appendix
\section{DM halo formation scales within the Press-Schechter paradigm}\label{sec:AppA}
%%%%%%%%%%%%%%%%%%%%%%%%%%%%%%%%%%%%%%%%%%%%%%%%%%%%

%Organization of the appendix: following on Arguelles-Diaz-...

%Why do we need boundary conditions: vital to the thermodynamic analysis
In the context of the calculation of solutions that represent fermionic DM halos via the MEPP, as shown in \cref{sec:II} and initially developed in \cite{2021MNRAS.502.4227A}, a key parameter comes from the boundary conditions for the solutions, for these general quasi-equilibrium solutions can only describe DM halos if their masses and radii correspond to the ones characteristic of such structures \cite{2018PDU....21...82A,2019PDU....24..278A}.
Indeed, the conclusions of such analysis regarding the thermodynamic stability of this system are susceptible to such contour conditions, with other authors reaching different conclusions by using the same analysis when considering systems smaller than galactic halos \cite{2015PhRvD..92l3527C,2020EPJB...93..208A}. Thus, appropriately choosing the mass and radius corresponding to a galactic halo becomes important. While DM halo masses are usually a well-estimated observable for most observed structures \cite{Karukes2019a}, defining the radius of such structures can depend on the particular cosmological model considered (see, e.g., \cite{MoBoschWhiteBook} for a thorough explanation). A common measure for such radii is the virial radius, obtained from applying the virial theorem to the gravitationally bound structure. 
However, due to differences arising from the particular collapse models used, it is common to use instead the $r_{200}$ radius, defined as the radius where the density of the system is $200$ times the background DM density \cite{BinneyTremaineBook}, which is close to the value of the virial radius $r_{\rm vir}$ for most models \cite{Bryan1997}. Thus, the relation between mass and $r_{200}$ radius for these structures is relatively straightforward:
\begin{equation}
M_{\rm vir} = 200 \frac{4}{3} \pi r_{200}^3 \rho_c (t) = 100 \frac{H_0^2 r_{200}^3 [1+z(t)]^3 }{G}  \ ,
\label{eq:SecPS_M200}
\end{equation}
where $M_{\rm vir}$ stands for the (virial) mass of the object. We see, however, that this definition is now time-dependent, as the background density of the universe evolves with time. In a cosmological context, we can also reinterpret this time dependence as the expansion of the spatial scales involved in the problem with time.

To marginalize this time dependence, a first approach can be to estimate the most likely collapse time of a structure of size $M_{\rm vir}$ and obtain the physical radius corresponding to such scale. While a deep study of this collapse time would typically involve a full study on nonlinear cosmology and structure formation (such as, e.g., \cite{Navarro1997,Maccio2012,Fitts2019}), it is sufficient for this study to use the Press-Schecter formalism \cite{PressSchechter74}, based on the results from linear cosmology for a given model. This is a well-studied theory, and a full description can be found, for example, in \cite{MoBoschWhiteBook,BinneyTremaineBook} among many other works. Here, we will limit ourselves to a summary of the assumptions and results. To study this formalism, it is necessary first to study how the overdensities collapse in the late, matter-dominated universe. For this, we can use one of the simpler nonlinear collapse models, the spherical collapse (see, e.g., \cite{MoBoschWhiteBook}). Under the assumptions of this model, any given overdensity collapses to form a virialized structure, and at that point, the linear theory of perturbation evolution will predict an overdensity of $\delta_c \sim 1.69$. So, according to this theory, we can assume that given an overdensity field evolving according to linear dynamics, a peak of density $> \delta_c$ would instead correspond to a collapsed halo. The core assumption of Press-Schechter's formalism regards how to relate this observation to the halo mass function. The formalism proposes that the fraction of the universe's total mass that is in halos of masses greater than $M$ is equal to the probability of a given overdensity is greater than the critical $\delta_c$, where the overdensity field $\delta_M$(t) is averaged with a filter of characteristic mass $M$. For this study, knowing the exact mass fraction of a given mass $M$ is not important. Instead, we are interested in the characteristic timescale at which overdensities of mass $M$ are most likely to collapse. In this case, we can assume most overdensities will start to collapse when the standard deviation $\sigma_M$ of the Gaussian field $\delta_M$ (known as the \textit{mass variance}) crosses the threshold of $\delta_c$. Thus, for an overdensity of mass $M$, we define a characteristic collapse time as 
\begin{equation}
\sigma_M (z^*) = \delta_c,
\label{eq:appxA_sigma_delta_relation}
\end{equation}
where
\begin{equation}
\sigma^2_M (t) = \frac{1}{2\pi^2} \int_0^{\infty} P(k) D^2(t) W^2(k,R) k^2 {\rm d}k,
\label{eq:appxA_sigma_definition}
\end{equation}  
being $D(t)$ the linear growth rate of perturbations, $P(k)$ the linear matter power spectrum, $W(k,R)$ a window function of characteristic radius $R(M)$ (taken as a \textit{top hat function} here, see e.g. \cite{MoBoschWhiteBook}), and $\delta_c \simeq 1.69$ according to spherical collapse \cite{MoBoschWhiteBook}.

We note that to obtain this characteristic collapse mass, it is necessary to know the linear matter power spectrum, typically obtained through a study on a particular cosmological model. In the preceding sections, we have seen a particle mass-dependent description of DM halo formation, where these masses are of order $\mathcal{O}$ 100 keV. This particle mass is significantly lighter than most CDM models suggest \cite{KolbTurner}, and instead points to an extension of the standard $\Lambda$CDM theory known as warm dark matter (WDM) \cite{Bode2000,Lovell2012,Boyarsky2018}. These types of extensions, typically characterized by DM particle masses in the keV ranges, show a considerably higher initial velocity dispersion when compared to their standard CDM counterparts, which in turn predicts a smaller number of small-scale structures in the universe and may alleviate some existing tensions with CDM \cite{Bullock2017}. The simplest production scenario for WDM is that these particles are created via thermal processes very early in the universe's history, and indeed this production scenario requires some degrees of freedom in the initial plasma well above $10^3$ \cite{MoBoschWhiteBook}. In any case, this scenario is an interesting benchmark for other production scenarios, as most result in a similar suppression feature in the matter power spectrum. An interesting class of models that can realize this WDM scenario can be found in sterile neutrino WDM, where non-equilibrium production of these particles can account for the observed DM fraction \cite{Boyarsky2018,Adhikari2016}, and particle self-interaction can reconcile the parameter space of these models with observation \cite{Yunis2021b}.

\begin{figure}
     %\centering
     \center{\includegraphics[width=\hsize]{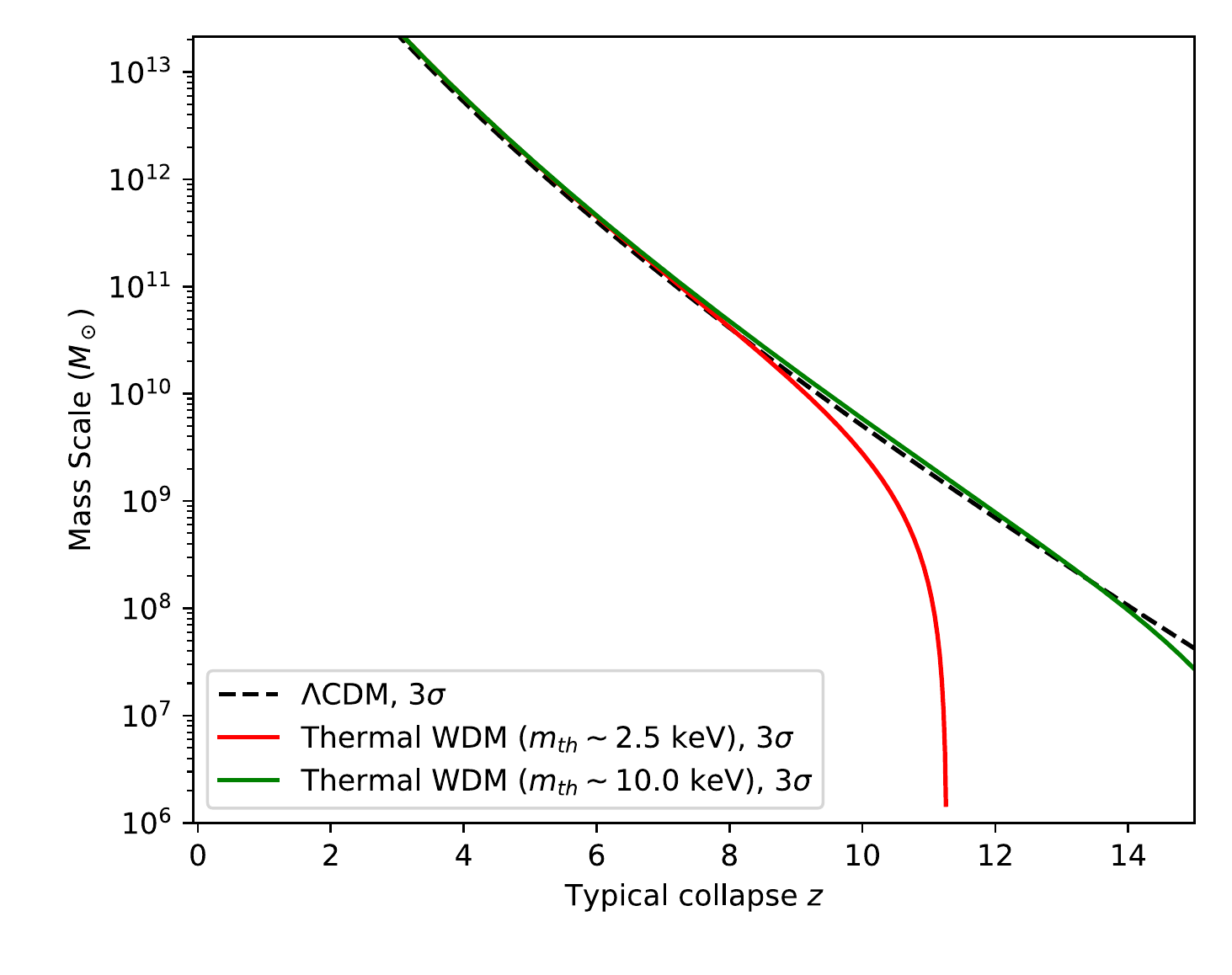}}
     %\includesvg[width=\columnwidth]{figs/ZvsM.svg}
     \caption{Characteristic collapse redshift $z^*$ as a function of substructure mass $M$. CDM models are represented in black lines, and WDM thermally produced relic models with $mc^2 = 2.5$ keV and $mc^2 = 10.0$ keV in red, blue, and green, respectively. The models considered here are taken at the $3\sigma$ threshold in the sense of equation \eqref{eq:appxA_sigma_definition}, as an estimation of the earliest possible halos.} 
    \label{fig:appxA_ZvsM}
\end{figure}

\begin{figure}
     %\centering
     \center{\includegraphics[width=\hsize]{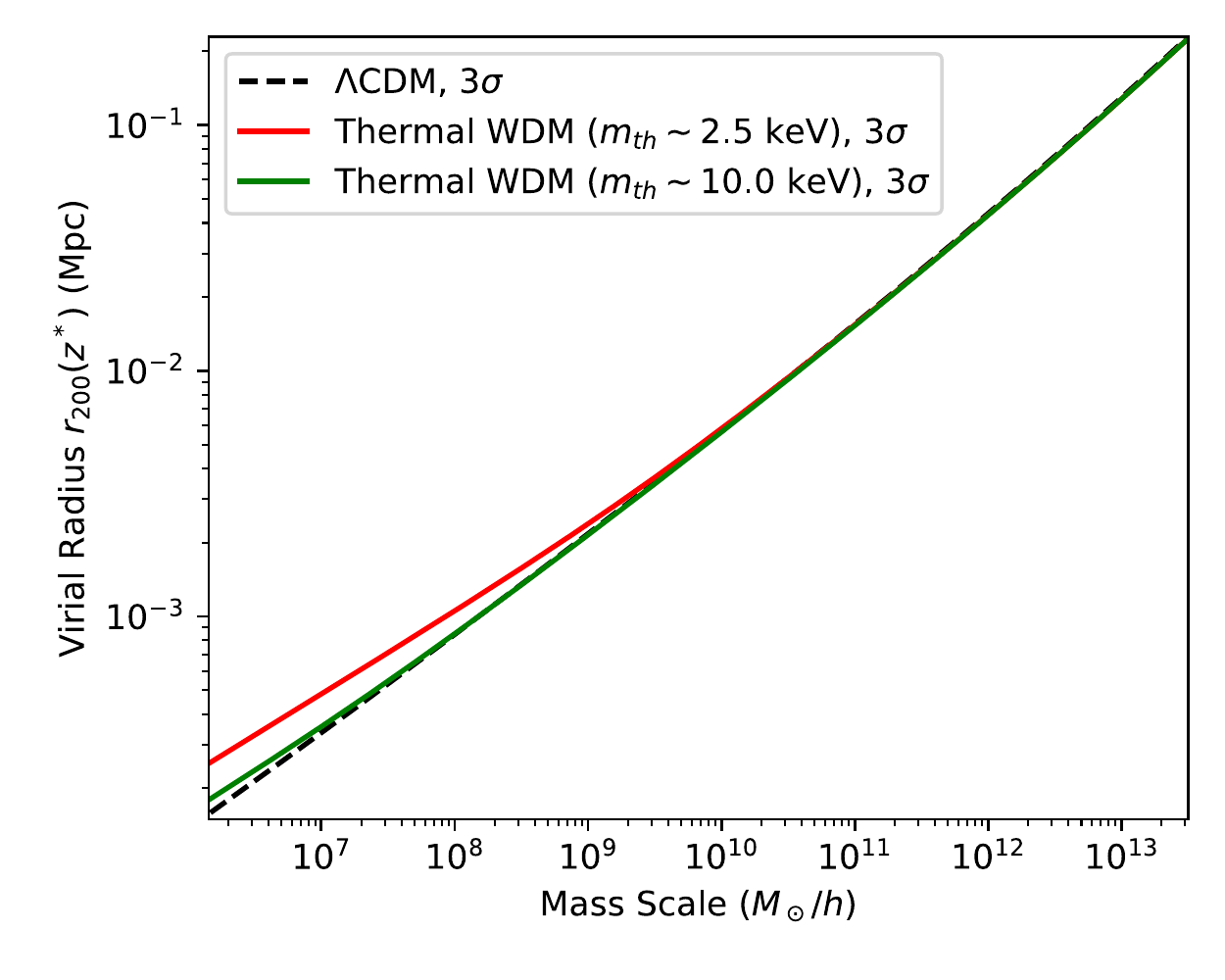}}
     %\includesvg[width=\columnwidth]{figs/MvsR200.svg}
     \caption{Virial radius $r_{200}$ at virialization time as a function of substructure mass $M$. CDM models are represented in black lines and WDM thermally produced relic models with $m c^2 = 2.5$ keV, $mc^2 = 5.0$ keV, and $mc^2 = 10.0$ keV in red, blue, and green, respectively. The models considered here are taken at the $3\sigma$ threshold in the sense of equation \eqref{eq:appxA_sigma_definition}, as an estimation of the earliest possible halos.} 
    \label{fig:appxA_MvsR200}
\end{figure}

% Show results
We can see the results for the $r_{200}$ as a function of the halo mass scale $M$ in figure \cref{fig:appxA_MvsR200} according to the Press-Schechter formalism, calculated for three thermally produced WDM models as well as a standard $\Lambda$CDM model. 
For reference, these $2.5$ and $10$ keV thermal models have similar suppression features as $\sim 15$ and $100$ keV non-resonantly produced sterile neutrinos, according to the criteria of \cite{Viel2013}.
We also include a line representing a CDM model, but where  \cref{eq:appxA_sigma_definition} has been refactored with $3\sigma$ instead to represent the time when earliest collapsed structures in the formalism are formed. There are thought to be produced roughly when rare $3\sigma$ overdensities cross the $\delta_c$ barrier and enter nonlinear collapse. However, in the mass scales relevant to these studies, the differences in $r_{200}$ between WDM and CDM models are minimal, and these cosmologies can be used interchangeably to calculate DM halo solutions.

%%%%%%%%%%%%%%%%%%%%%%%%%%%%%%%%%%%%%%%%%%%%%%%%%%%%
\section{Geodesic disk accretion}\label{app:B}
%%%%%%%%%%%%%%%%%%%%%%%%%%%%%%%%%%%%%%%%%%%%%%%%%%%%

We here present the equations of the evolution of a Kerr BH during the accretion of matter from a geodesic thin disk. The treatment closely follows the formulation in \cite{1973blho.conf..343N, 1974ApJ...191..499P, 1974ApJ...191..507T}. 

In the equatorial plane ($\theta=\pi/2$) and close above it, the Kerr spacetime metric can be written as
\begin{equation}
    ds^2 = - e^{2\nu} dt^2 + e^{2\psi} (d\phi-\omega dt)^2 + e^{2 \mu} dr^2 + dz^2,
\end{equation}
where $z$ is the height above the equatorial plane, and $\nu$, $\psi$, $\mu$, $\omega$ are functions of the radial coordinate $r$:
\begin{equation}\label{eq:metric}
    e^{2\nu} = \frac{r^2\Delta}{A},\quad e^{2\psi} = \frac{A}{r^2},\quad e^{2\mu} = \frac{r^2}{\Delta}, \quad
    \omega = \frac{2 M a r}{A},
\end{equation}
being $\Delta=r^2-2 M r+ a^2$, and $A = (r^2+a^2)^2-\Delta a^2$, with $M$ and $a=J/M$, respectively, the BH mass and angular momentum per unit mass.

\begin{figure*}
\begin{minipage}{0.49\linewidth}
\center{\includegraphics[width=0.97\linewidth]{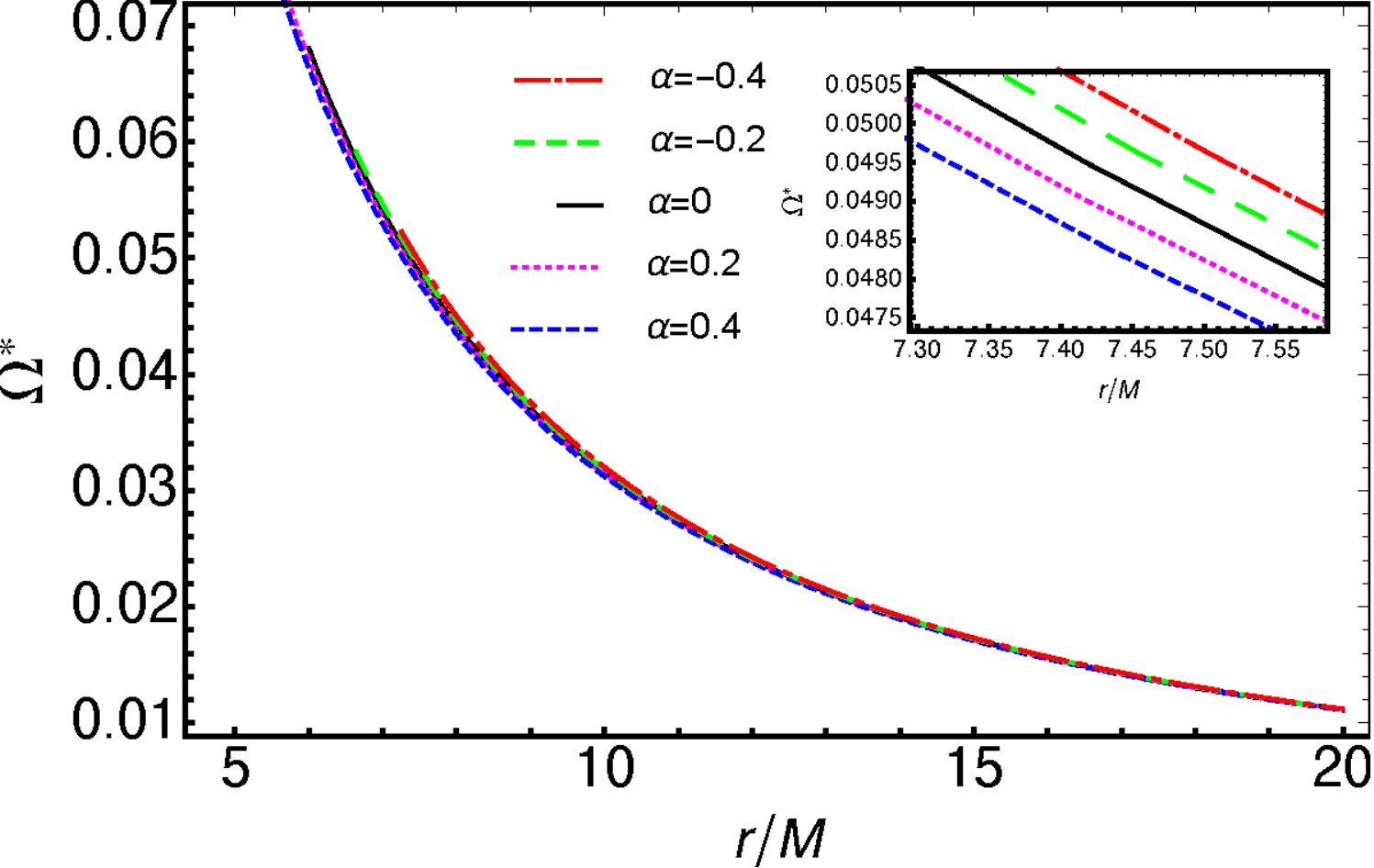}\\ }
\end{minipage}
\hfill 
\begin{minipage}{0.50\linewidth}
\center{\includegraphics[width=0.97\linewidth]{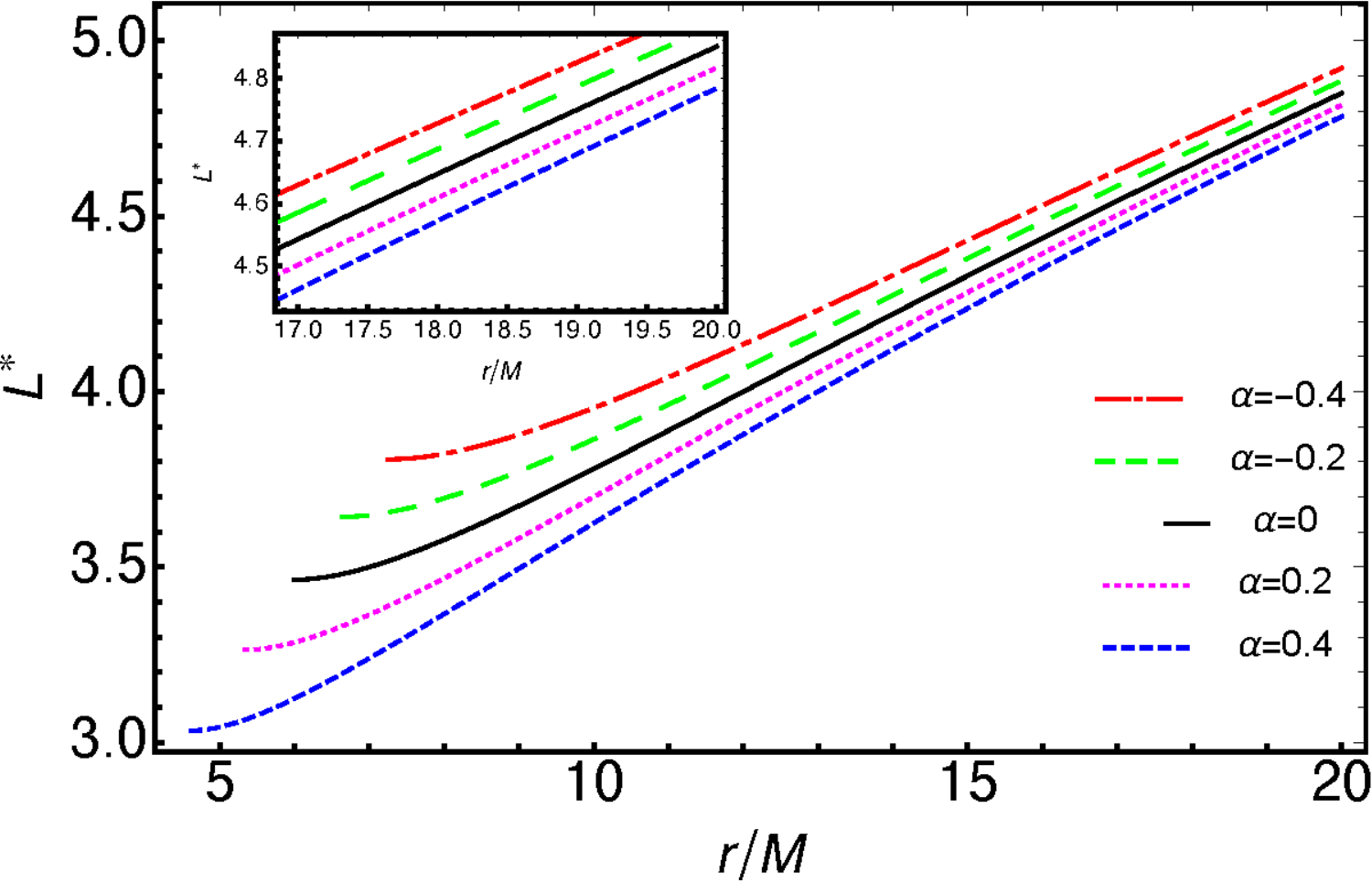}\\ }
\end{minipage}
\caption{Left panel: Angular velocity of test particles versus radial distance $r$ in units of the BH mass, $M$. Right panel: Angular momentum $L^*$ of test particles versus radial distance $r$ in units of the BH mass, $M$}
\label{fig:angvelmom}
\end{figure*}

\begin{figure*}%[ht]
\begin{minipage}{0.49\linewidth}
\center{\includegraphics[width=0.97\linewidth]{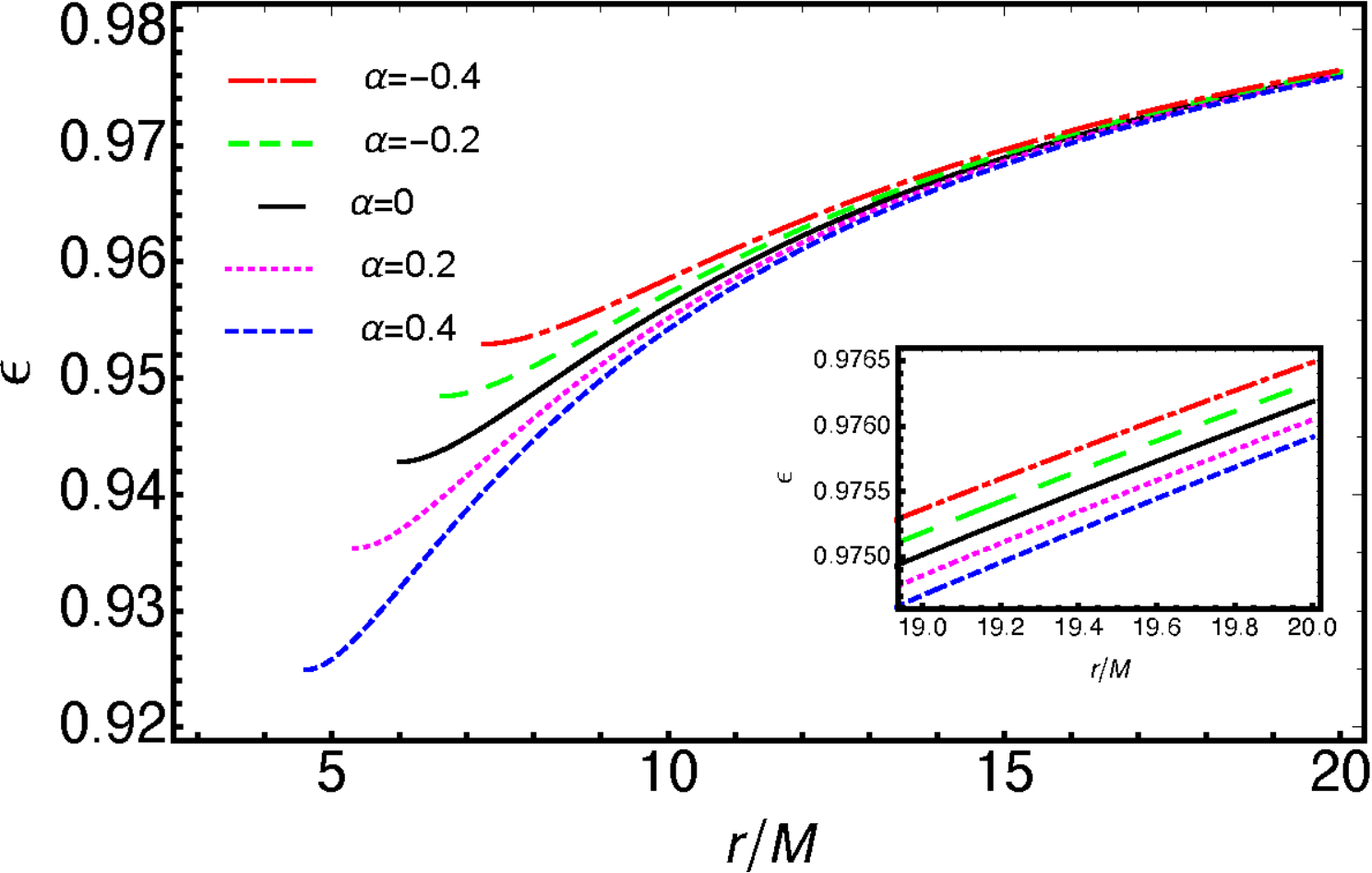}\\ }
\end{minipage}
\hfill 
\begin{minipage}{0.50\linewidth}
\center{\includegraphics[width=0.97\linewidth]{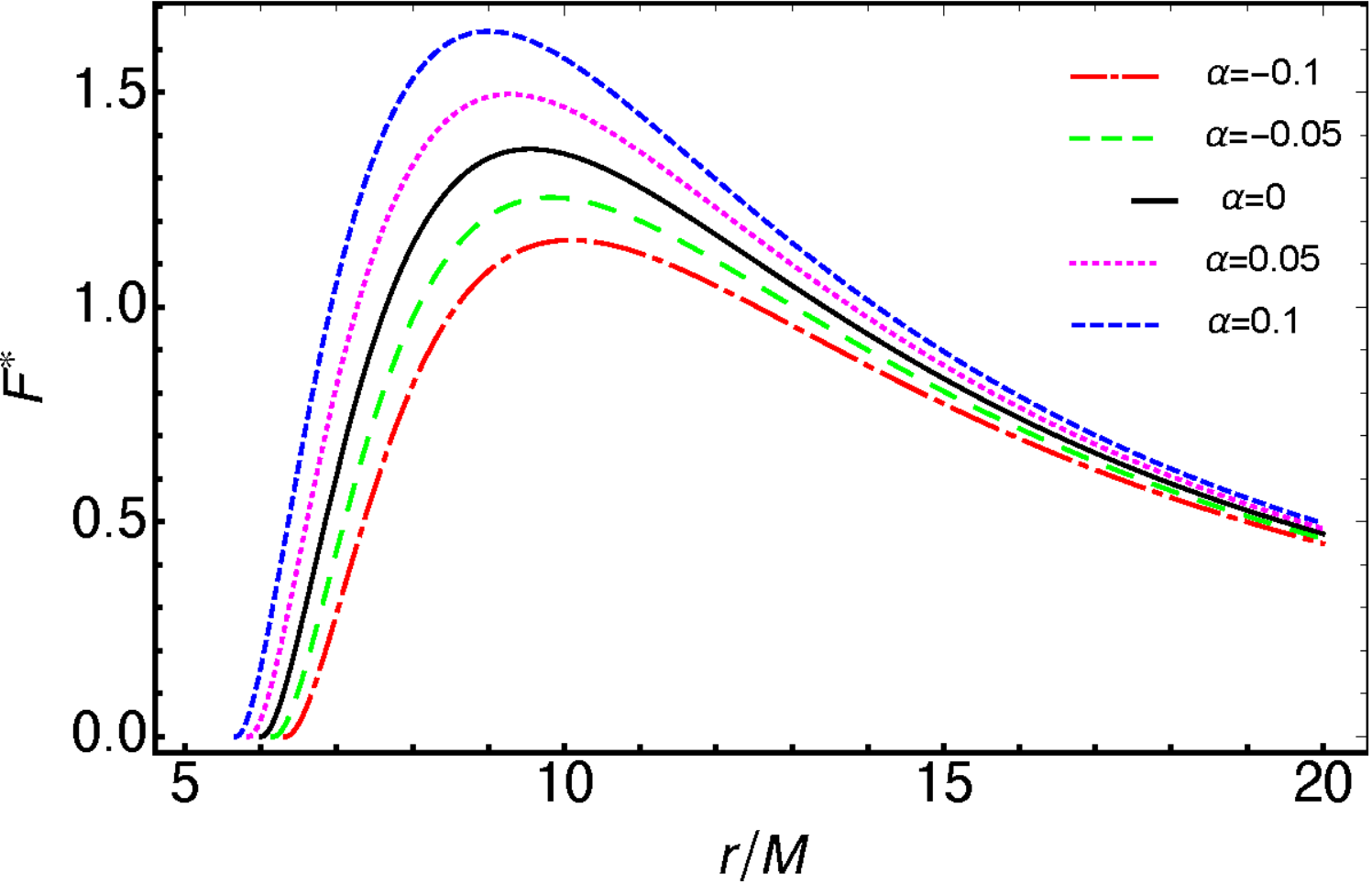}\\ }
\end{minipage}
\caption{Left panel: Energy $E^*$ of test particles versus radial distance $r$ in units of the BH mass, $M$. Right panel: radiative flux $\mathcal{F}^*$ multiplied by $10^5$ of the accretion disk versus radial distance $r$ in units of the BH mass, $M$.}
\label{fig:energyflux}
\end{figure*}

Matter and radiation transfer energy and angular momentum to the BH during the accretion. In particular, it is essential to account for the feedback of radiation/photons onto the BH since they exert a counter-torque \cite{1970PhRvD...1.2721G} that avoids the BH from reaching the extreme regime $a=M$. This implies that the accretion of massive particles and radiation does not lead the BH to become a naked singularity \cite{1974ApJ...191..507T}. We denote by $dm$ the rest-mass accreted by the BH in a coordinate time interval $dt$, so $\dot{m} = dm/dt$ is the rest-mass accretion rate, and $\dot{M}_{\rm rad}$ and $\dot{J}_{\rm rad}$ are the rate of energy and angular momentum transfer by radiation to the BH. Therefore, we can write the BH energy and angular momentum conservation equations as \cite{1974ApJ...191..507T}
\begin{align}
    \dot{M} &= \dot{M}_{\rm matter} + \dot{M}_{\rm rad},\label{eq:Mdottotal}\\
    \dot{J} &= \dot{J}_{\rm matter} + \dot{J}_{\rm rad},\label{eq:Jdottotal}
\end{align}
where 
\begin{align}
    \dot{M}_{\rm matter} &= \epsilon_0\,\dot{m},\label{eq:Mdotmatter}\\
    \dot{J}_{\rm matter} &= l_0\,\dot{m},\label{eq:Jdotmatter}\\
    \dot{M}_{\rm rad} &= -\frac{2}{\pi}\int_{r_{0}}^{\infty}\!\int_{0}^{\pi/2}\!\!\int_ {0}^{2\pi}\!\! C k_t F(r) dS, \label{eq:Mdotrad}\\
    \dot{J}_{\rm rad} &= \frac{2}{\pi}\int_{r_{0}}^{\infty}\!\int_{0}^{\pi/2}\!\!\int_ {0}^{2\pi}\!\! C k_\phi F(r) dS,\label{eq:Jdotrad}
\end{align}
being $\epsilon_0$ and $l_0$ the specific (i.e., per unit mass) energy and angular momentum of the matter accreted, i.e., at the radius $r_0$ of the innermost stable circular orbit (ISCO). Assuming that radiation emitted from the disk's surface is isotropic, the normalized photon four-momentum, as measured by a comoving observer, is $k^{\tilde{\mu}} = p^{\tilde{\mu}}/p^{\tilde{0}} = (1,\sin\Theta \cos\Phi, \sin\Theta \sin\Phi, \cos\Theta)$ and $k^{\mu} = p^\mu/p^{\tilde{0}}$, where $p^\mu$ is the normalized photon four-momentum in the coordinate frame. The expressions for $k^{\mu}$ can be found in Appendix A of \cite{1974ApJ...191..507T} (see also Appendix C in \cite{2022ApJ...929...56R}).  The factor $C=C(r,\Theta,\Phi)$ is the \textit{capture function} that takes the value $1$ or $0$ when photons emitted at radius a $r$ with local direction $(\Theta,\Phi)$ are captured by the BH or escape to infinity, respectively. The surface area element is $dS = 2 \pi r \sin\Theta \cos\Theta d\Phi d\Theta dr$. The function $F(r)$ is the radiation flux emitted from the disk measured by the comoving observer \cite{1973blho.conf..343N, 1974ApJ...191..499P}
\begin{equation}\label{eq:F}
    F(r) = -\frac{\dot{m}}{4\pi \sqrt{-g}} \frac{\Omega_{,r}}{(\epsilon-\Omega l)^2}\int_{r_0}^r (\epsilon-\Omega l) l_{,\bar{r}} d\bar{r},
\end{equation}
where $\sqrt{-g} = e^{\nu+\psi+\mu} = r$, $\epsilon$, and $l$ is the specific energy and angular momentum of circular geodesics of radius $r$ in Kerr-metric (measured at infinity), and $\Omega = u^\phi/u^t$ is their angular velocity measured in the coordinate frame, being $u^\mu$ the fluid four-velocity \cite{1973blho.conf..215B}. Using the change of variable $x=\sqrt{r/M}$ we can write
\begin{align}
    \epsilon &= \frac{x^3-2x\pm\alpha}{x^{3/2}\sqrt{x^3-3x\pm 2\alpha}},\label{eq:epsisco}\\
    l &= \pm \frac{M\left(x^4 \mp 2\alpha x +\alpha^2\right)}{x^{3/2}\sqrt{x^3-3x \pm 2\alpha}},\label{eq:lisco}\\
    \Omega &= \frac{1}{M}\frac{1}{\alpha \pm x^3},\label{eq:Omega}
\end{align}
where $\alpha = a/M$ is the dimensionless spin parameter. The upper/lower sign corresponds to co-rotating/counter-rotating circular orbits. Clearly, we have $\epsilon_0 = \epsilon(r_0)$ and $l_0 = l (r_0)$. The radius of the ISCO is given by \cite{1973blho.conf..215B}
\begin{align}
    r_0 &= M \left[ 3 + Z_2 \mp  \sqrt{(3-Z_1)(3+Z_1+2 Z_2)} \right],\\
    Z_1 &= 1 + (1-\alpha^2)^{1/3} \left[(1+\alpha)^{1/3} + (1-\alpha)^{1/3} \right], \\
    Z_2 &= \sqrt{3 \alpha^2 + Z_1},
\end{align}

In \cref{fig:angvelmom}, we plot the orbital angular velocity $\Omega^*=M \Omega$ (left panel) and angular momentum $L^*=l/M$ (right panel) of test particles as a function of radial coordinate in the Kerr metric, for selected values of the spin parameter. Counter-rotating test particles possess larger angular velocity and momentum than co-rotating particles (for details see \cite{2020MNRAS.496.1115B,2021PhRvD.104h4009B}).

In \cref{fig:energyflux}, the energy per unit mass $\epsilon$ of test particles is shown as a function of radial coordinate (left panel) and the radiative flux $F^*=10^5M^2F/\dot{m}$ emitted from the accretion disk as a function of radial coordinate (right panel) for different values of the spin parameter.

%\FloatBarrier

%%%%%%%%%%%%%%%%%%%%%%%%%%%%%%%%%%%%%%
% \begin{acknowledgments}
% ...
% \end{acknowledgments}

\end{document}